\documentclass[12pt]{article}
\usepackage{graphicx}
\usepackage{lscape}
\usepackage{citesort}

\begin{document}

\def\ds{\displaystyle}
\def\beq{\begin{equation}}
\def\eeq{\end{equation}}
\def\bea{\begin{eqnarray}}
\def\eea{\end{eqnarray}}
\def\beeq{\begin{eqnarray}}
\def\eeeq{\end{eqnarray}}
\def\ve{\vert}
\def\vel{\left|}
\def\ver{\right|}
\def\nnb{\nonumber}
\def\ga{\left(}
\def\dr{\right)}
\def\aga{\left\{}
\def\adr{\right\}}
\def\lla{\left<}
\def\rra{\right>}
\def\rar{\rightarrow}
\def\nnb{\nonumber}
\def\la{\langle}
\def\ra{\rangle}
\def\ba{\begin{array}}
\def\ea{\end{array}}
\def\qs{\la \bar s s \ra}
\def\qu{\la \bar u u \ra}
\def\qd{\la \bar d d \ra}
\def\qq{\la \bar q q \ra}
\def\gGgG{\la g^2 G^2 \ra}
\def\es{\!\!\! &=& \!\!\!}
\def\ap{\!\!\! &\approx& \!\!\!}
\def\ar{&+& \!\!\!}
\def\ek{&-& \!\!\!}
\def\kek{\!\!\!&-& \!\!\!}
\def\cp{&\times& \!\!\!}
\def\se{\!\!\! &\simeq& \!\!\!}


\def\simlt{\stackrel{<}{{}_\sim}}
\def\simgt{\stackrel{>}{{}_\sim}}


\title{{\Large {\bf
QCD sum rules analysis of the rare  $B_c \rar X\nu\bar{\nu}$ decays
}}}

\author{\vspace{1cm}\\
{\small K. Azizi \thanks {e-mail: e146342@metu.edu.tr}~\,} \\
{\small  Department of Physics, Middle East Technical University,
06531 Ankara, Turkey}\\
{\small R. Khosravi \thanks {e-mail: khosravi.reza @ gmail.com}~\,}\\
 {\small Physics Department , Shiraz University, Shiraz 71454,
Iran}\\
{\small V. Bashiry \thanks {e-mail: bashiry@ciu.edu.tr}~\,}\\
 {\small Electric-Electronics Department, Cyprus International University, Via Mersin 10 ,
 Turkey}\\}
\date{}

\begin{titlepage}
\maketitle
\thispagestyle{empty}

\begin{abstract}
 Taking into account
 the gluon correction contributions to the correlation function,  the form factors relevant to the  rare  $B_c \rar X \nu\bar{\nu}$ decays
  are
 calculated in the framework of the three point QCD sum rules, where $X$ stands for   axial vector particle, $AV(D_{s1})$, and vector particles, $V(D^*,D^*_s)$. The total decay width as well as the branching ratio of these decays are evaluated using
 the $q^2$ dependent  expressions of the form factors. A comparison of our
 results with the predictions of the relativistic constituent quark
 model is presented.
\end{abstract}

\end{titlepage}

\section{Introduction}
The discovery of the $B_{c}$ meson by the CDF detector at the Fermi
Lab  in $p\bar p$ collisions via the decay mode $Bc \rightarrow
J/\psi l^{\pm} \nu$ at $\sqrt{s} = 1.8 ~TeV$ \cite{Abe} has
 illustrated  the possibility of the experimental study of the
charm-beauty systems and has produced  considerable interest in its
spectroscopy. This meson constitutes a very rich laboratory since
with the luminosity values of ${\cal L}=10^{34}cm^{-2}s^{-1}$ and
$\sqrt{s}=14\rm TeV$ at LHC, the number of $B_c^{\pm}$ mesons is
expected to be about $10^{8}\sim10^{10}$ per year \cite{Du,Stone}.
This will provide a good opportunity  to study not only some rare
$B_c$ decays, but also CP violation, T violation and polarization
asymmetries. The long-lived heavy quarkonium, $B_{c}$, is the only
meson containing two heavy quarks with different charge and flavours
(b and c), in which its decay properties are expected to be
different from that of flavour neutral mesons and this can produce a
significant progress in the study of heavy quark dynamics. The
$B_{c}$ system is the lowest bound state of two heavy quarks with
open flavour. Such states have no annihilation decay modes due to
the electromagnetic and strong interactions since the excited levels
of $\bar b c$ lie below the threshold of decay into the pair of
heavy B and D mesons, so this meson decays weakly. Many parameters
enter in the description of weak decays of this meson. In
particular, measuring the branching ratios of such decays provide a
new framework for more precise calculation of the
Cabibbo-Kobayashi-Maskawa (CKM) matrix elements $V_{tq}$ (q = d, s,
b), leptonic decay constants, quark masses and mixing angles. Also,
a study of this meson can be used as constrains  on the physics
beyond the Standard Model. Indeed, more collection of hadrons
containing heavy quarks provides more accuracy and confidence in the
understanding of QCD dynamics (for details about the physics of the
$B_{c}$ meson see \cite{Gershtein}).

In present work, the $B_c \rar (D^*,D^*_s) \nu\bar{\nu}$ and $B_c
\rar D_{s1} \nu\bar{\nu}$ transitions are investigated in
the framework of the three point QCD sum rules. Theoretical
calculation of the  amplitudes for these decays is particularly
reliable, owing to the absence of long-distance interactions that
affect charged-lepton channels $B_c \rar X l^{^+}l^{-}$. The rare
$B_c \rar (D^*,D^*_s) \nu\bar{\nu}$ and  $B_c \rar D_{s1}
\nu\bar{\nu}$
 decays are proceeded by flavour changing neutral current (FCNC)
transitions of $b \rightarrow s, d$. These transitions occur at loop
level in the standard model (SM) and they are very sensitive to the
physics beyond the SM, since some new particles might have
contributions in the loops diagrams. New physics such as SUSY
particles or a possible fourth generation could contribute to the
penguin loop or box diagram and change the branching fractions
\cite{Buchalla}. The possibility of discovering light dark matter in
$b \rightarrow s$ transitions with large missing momentum has been
discussed in Ref.~\cite{Bird}. Note that, some possible $B_{c}$
decays such as $B_{c}\rightarrow l \overline{\nu}\gamma$,
$B_{c}\rightarrow \rho^{+}\gamma$, $B_{c}\rightarrow
K^{\ast+}\gamma$, $B_{c}\rightarrow B_{u}^{\ast}l^{+}l^{-}$,
$B_{c}\rightarrow B_{u}^{\ast}\gamma $ and $B_{c}\rightarrow
D_{s,d}^{\ast}\gamma $
 have been previously studied in the framework of light-cone  and three
point QCD sum rules \cite{Aliev1,Aliev2,Aliev3,Alievsp,azizi1}. A
larger set of exclusive nonleptonic and semileptonic decays of the
$B_{c}$ meson, which have been studied within a relativistic
constituent quark model can be found in Ref.~\cite{Ivanov}.
Moreover, the $B_c \rar (D^*,D^*_s) \nu\bar{\nu}$ transitions have
also been investigated in the framework of the relativistic
constituent quark
 model (RCQM)
\cite{tt1}.

The paper includes three sections. The calculation of the sum
rules for the relevant form factors are presented in section II.
In the sum rules expressions for the form factors, the light quark
condensates don't have any contributions, so, as first correction
in the nonperturbative part of the correlator,  the two gluon
condensates contributions to the correlation function are taken
into account. Section III contains numerical analysis, discussion
and comparison of the present work results with the predictions of
the RCQM.
\begin{figure}
\vspace*{-1cm}
\begin{center}
\includegraphics[width=11cm]{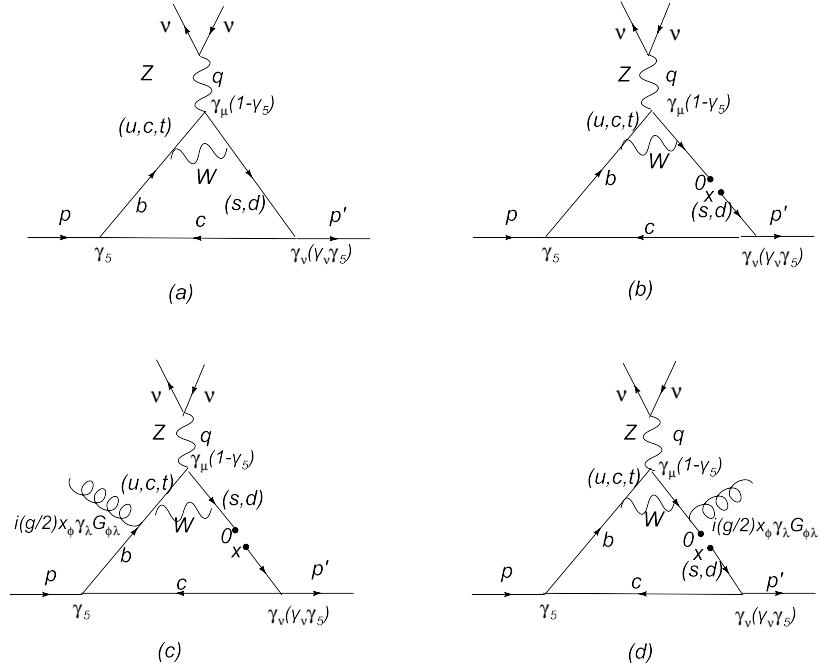}
\end{center}
\caption{loop diagrams for $B_c \rar X \nu\bar{\nu}$ transitions,
bare loop (diagram a) and light quark condensates (without any gluon
diagram b and with one gluon emission diagrams c, d)} \label{fig1}
\end{figure}
\section{Sum rules for transition form factors of $B \rightarrow AV (V)\nu \bar \nu$ $[AV(V)=D_{s1}(D^*,D^*_s)]$}
 The  $B \rightarrow X \nu \bar \nu$  process is
described at quark level via the $b \rightarrow q \nu \bar \nu$
transition,$(q=d$ or $ s)$ (see Fig. 1), in the SM and receives
contributions from $Z$-penguin and box diagrams, where dominant
contributions come from intermediate top quark. The explicit form of
the effective Hamiltonian responsible for $b \rightarrow q \nu \bar
\nu$ decays is described by only one Wilson coefficient, namely
$C_{10}$,
\begin{equation}\label{1abu}
H_{eff} =\frac{G_{_F} \alpha}{2 \pi \sqrt {2}}~ C_{10}
~(V_{tb}V_{tq}^*)~ \overline{q}{\gamma}^{\mu}(1-{\gamma}_5)b~
\overline{\nu}{\gamma}_{\mu}(1-{\gamma}_5)\nu , \label{(1)}
\end{equation}
where $G_{_F}$ is the Fermi constant, $\alpha$ is the fine structure
constant at the $Z$ mass scale and $V_{ij}$ are elements of the CKM
matrix.  The presence of only one operator in
the effective Hamiltonian makes the  $ b \rightarrow q\nu\bar\nu$
process important, because the estimated theoretical uncertainty is
related only to the value of the Wilson coefficient $C_{10}$. For
more about the Wilson coefficients see \cite{Aliev3,tt1,ref16,ref15} and
references therein. The amplitudes of the $B_{c}\rightarrow
X\nu\bar{\nu}$  decays are obtained by  sandwiching
Eq. (1) between the initial and final meson states
\begin{equation}\label{2au}
M=\frac{G_{_F} \alpha}{2 \pi \sqrt {2}}~ C_{10}
~(V_{tb}V_{tq}^*)~\overline{\nu}
~\gamma_{\mu}(1-\gamma_{5})\nu<X(p',\varepsilon)\mid~\overline{q}
~\gamma_{\mu}(1-\gamma_{5}) b\mid B_{c}(p)>.
\end{equation}

Our aim is to  calculate   the matrix element
$<X(p',\varepsilon)\mid\overline{q}\gamma_{\mu}(1-\gamma_{5}) b\mid
B_{c}(p)>$ appearing in Eq. (\ref{2au}). Both vector and axial
vector part of the transition current
  $~\overline{q}~\gamma_{\mu}(1-\gamma_{5}) b~$  contribute to the
matrix element discussed above. Considering the parity and Lorentz
invariances, one can parameterized this matrix element  in terms
of the form factors in the following form:

\begin{equation}\label{3au}
<V(p',\varepsilon)\mid\overline{q}\gamma_{\mu} b\mid
B_c(p)>=-\epsilon_{\mu\nu\alpha\beta} \varepsilon^{\ast\nu}
p^\alpha p^{\prime\beta} \frac{2 V(q^2)}{m_{B_c} + m_V},
\end{equation}

\bea \label{4au} \lla V (p^\prime,\varepsilon) \vel \bar{q}
\gamma_\mu \gamma_{5} b \ver B_c (p)\rra &=& - i \Bigg[
\varepsilon_\mu^\ast (m_{B_c} + m_{V}) A_1(q^2) -(\varepsilon^\ast
q) {\cal P}_\mu \frac{A_2(q^2)}{m_{B_c} +
m_{V}}\nnb \\
&-& (\varepsilon^\ast q) \frac{2 m_{V}}{q^2} [A_3(q^2) -A_0(q^2)]
q_\mu \Bigg]~. \eea
\begin{equation}\label{300au}
<AV(p',\varepsilon)\mid\overline{q}\gamma_{\mu}\gamma_{5} b\mid
B_c(p)>=-\epsilon_{\mu\nu\alpha\beta} \varepsilon^{\ast\nu} p^\alpha
p^{\prime\beta} \frac{2 V'(q^2)}{m_{B_c} + m_{AV}},
\end{equation}

\bea \label{400au} \lla AV (p^\prime,\varepsilon) \vel \bar{q}
\gamma_\mu  b \ver B_c (p)\rra &=& - i \Bigg[ \varepsilon_\mu^\ast
(m_{B_c} + m_{AV}) A'_1(q^2) -(\varepsilon^\ast q) {\cal P}_\mu
\frac{A'_2(q^2)}{m_{B_c} +
m_{AV}}\nnb \\
&-& (\varepsilon^\ast q) \frac{2 m_{AV}}{q^2} [A'_3(q^2) -A'_0(q^2)]
q_\mu \Bigg]~. \eea
 where ${\cal P}_\mu =
(p+p^\prime)_\mu$, $q_\mu=(p-p^\prime)_\mu$ and $\varepsilon^\ast$
is the polarization vector of the  $X$ mesons. To guarantee the
finiteness of the results at $q^{2} = 0$, it should be
$A_{3}(0)(A'_{3}(0)) = A_{0}(0)(A'_{0}(0))$. The form factor
$A_3(q^2)(A'_3(q^2))$  can be written as a linear combination of
$A_1(A'_1)$ and  $A_2(A'_2)$ in the following way:

\bea \label{e7306} A_3(q^2)(A'_3(q^2)) = \frac{m_{B_c} +
m_{V}(m_{AV})}{2 m_{V}(m_{AV})} A_1(q^2)(A'_1(q^2)) - \frac{m_{B_c}
- m_{V}(m_{AV})}{2 m_{V}(m_{AV})} A_2(q^2)(A'_2(q^2))~. \eea

Therefore, we need to calculate the form factors $V(V')$,
$A_1(A'_1)$ and $A_2(A'_2)$. In order to obtain the sum rules
expressions for these form factors,  we consider   the following
three-point correlation function: \bea \label{e7308}
\Pi_{\mu\nu}^{v;a} = i^2\int d^{4}xd^4ye^{-ipx}e^{ip'y}\la 0 \ve
{\cal T} \left\{ J_{X \nu}(y) J_\mu(0)^{v;a}  J^{\dag}_{B_c}(x)
\right\} \ve 0 \ra~, \eea
\\
where $J _{V \nu }(y)=\overline{c}\gamma_{\nu} q$, $J _{AV \nu
}(y)=\overline{c}\gamma_{5}\gamma_{\nu} q$  $(q= s,d)$ and
$J_{B_{c}}(x)=i\overline{c}\gamma_{5}b$ are the interpolating
current of the  $ V $,  $ AV $ and  $B_{c} $ mesons, respectively.
$J_{\mu}^{v}=~\overline{q}\gamma_{\mu}b $ and
$J_{\mu}^{a}=~\overline{q}\gamma_{\mu}\gamma_{5}b$ are the
 vector and axial vector part of the transition current.

From the general philosophy of the QCD sum rules, we calculate the
above correlation function in two languages. First, in the hadron
languages, the  results of the correlator give us the
phenomenological or physical part and the QCD or theoretical part
of this correlator are obtained in the quark gluon languages. The
sum rules for the form factors can be obtained by equating the
coefficient of the  corresponding  structure from these two parts
and applying double Borel transformation with respect to the
momentums of the initial and final meson states to eliminate the
contributions coming from the higher states and continuum.

To calculate the phenomenological part of the correlator given in
Eq. (\ref{e7308}), two complete sets of intermediate states with the
same quantum numbers as the currents $J_{V}$ and $J_{B_{c}}$
 are inserted, respectively. As a result of this procedure, we get the
following representation of the above-mentioned correlator:
\begin{eqnarray} \label{7au}
&&\Pi _{\mu\nu}^{v;a}(p^2,p'^2,q^2)=
\nonumber \\
&& \frac{<0\mid J_{X \nu} \mid X(p',\varepsilon)><
X(p',\varepsilon)\mid J_{\mu}^{v;a}\mid B_{c}(p)><B_{c}(p)\mid
J^{\dag}_{Bc}\mid0>}{(p'^2-m_{X}^2)(p^2-m_{Bc}^2)}+\cdots
\nonumber \\
\end{eqnarray}
 where $\cdots$ represent contributions coming from higher states and continuum. The matrix
 elements in Eq. (\ref{7au}) are defined in the standard way as:
\begin{equation}\label{8au}
 <0\mid J^{\nu}_{X} \mid
X(p')>=f_{X}m_{X}\varepsilon^{\nu}~,~~<0\mid J_{Bc}\mid
B_{c}(p)>=i\frac{f_{B_{c}}m_{B_{c}}^2}{m_{b}+m_{c}},
\end{equation}
where $f_{X}$ and $f_{B_{c}}$  are the leptonic decay constants of
$X $ and $B_{c}$ mesons, respectively. Using Eqs.
(\ref{3au}-\ref{400au}) and Eq. (\ref{8au}) in Eq. (\ref{7au}) and
performing summation over the polarization of $X$ meson , we get the
following result for the physical part: \bea \label{e7312}
\Pi_{\nu\mu}^{(v;a)}(p^2,p'^2,q^2) \es  \frac{f_{B_c} f_{V(AV)}
m_{B_c}^2 m_{V(AV)}} {(m_b+m_c) (p_1^2 - m_{B_c}^2 ) ( p_2^2 -
m_{V(AV)}^2)} \Bigg\{i \epsilon_{\nu\mu\alpha\beta} p^\alpha
p^{\prime\beta} \frac{2
V(V')}{m_{B_c} + m_{V(AV)} } \nnb \\
\ek (m_{B_c} + m_{V(AV)}) \Bigg( -g_{\mu\nu} + \frac{
({\cal P}-q)_\mu ({\cal P}-q)_\nu}{4 m_{V(AV)}^2} \Bigg) A_1(A'_1) \nnb \\
\ar \frac{1}{m_{B_c} + m_{V(AV)}} {\cal P}_\mu \Bigg( - q_\nu +
\frac{p^\prime q ({\cal P}-q)_\nu}{2 m_{V(AV)}^2} \Bigg) A_2(A'_2) \nnb \\
\ar \frac{2  m_{V(AV)}}{q^2} q_\mu \Bigg( - q_\nu + \frac{p^\prime q
({\cal P}-q)_\nu}{2 m_{V(AV)}^2} \Bigg) (A_3(A'_3) - A_0(A'_0))
\Bigg\}~. \eea

The coefficients of the Lorentz structures
$i\epsilon_{\nu\mu\alpha\beta} p^\alpha p^{\prime\beta}$, $
g_{\mu\nu}$ and ${\cal P}_\mu q_\nu$ give the  expressions for  the
form factors $V(V')$, $A_1(A'_1)$ and  $A_2(A'_2)$, respectively.
The correlation function   can be written in trms of the Lorentz
structures in the following form: \bea \label{e7309}
\Pi_{\nu\mu}^{(v;a)}(p^2,p'^2,q^2) \es
\Pi_V\epsilon_{\nu\mu\alpha\beta} \,p^\alpha p^{\prime\beta}  +
\Pi_{A_1} g_{\mu\nu} + \Pi_{A_2} {\cal P}_\nu q_\mu +...\eea

To calculate the QCD side of correlation function, on the other
side, we evaluate the  three--point correlator by the help of the
operator product expansion (OPE) in the deep Euclidean region $p^2
\ll (m_b + m_c)^2$, $p'^2 \ll (m_b^2+m_{q}^2)$. For this aim, we
write each $\Pi_{i(i')}$ [$i(i')$ stands for $V(V')$,
$A_{1}(A'_{1})$ and $A_{2}(A'_{2})$] function in terms of the
perturbative and nonperturbative parts as:

\bea \label{e7316} \Pi_{i(i')}(p^2,p'^2,q^2) =
\Pi_{i(i')}^{per}(p_1^2,p_2^2,q^2)
+\Pi_{i(i')}^{non-per}(p^2,p'^2,q^2)~, \eea where   $<\bar qq>$ and
$<G^2>$ denotes the light quark and two gluon condensates,
respectively. For the perturbative part, we calculate the bare loop
diagram (Fig. 1 a), however, diagrams b, c, d in Fig. 1 are
correspond to the light quark condensates contributing to the
correlation function.  In principle, the light quark condensate
diagrams  give contributions to the correlation function, but
applying double Borel transformations kill their contributions, so
as first nonperturbative correction, we   consider  the gluon
condensate diagrams (see Fig. 2 a, b, c, d, e, f).

Using the double dispersion representation, the bare--loop
contribution is written as \bea \label{e7317} \Pi_{i(i')}^{per} = -
\frac{1}{(2 \pi)^2} \int ds^\prime \int ds \frac{\rho_{i(i')}^{per}
(s,s^\prime, Q^2)}{(s-p^2) (s^\prime - p^{\prime 2})}   + \mbox{\rm
subtraction terms}~, \eea  where $Q^2=-q^2$. The spectral densities
$\rho_{i(i')}^{per} (s,s^\prime,Q^2)$are calculated by  the help of
the  Gutkovsky rule, i.e., we replace the propagators with
Dirac--delta functions \bea \label{e7319} \frac{1}{p^2-m^2} \rar
-2i\pi \delta(p^2-m^2)~,\eea implying that all quarks are real and
integration region in Eq. (\ref{e7317}) is obtained by requiring
that the argument of three delta vanish, simultaneously. This
condition leads to the following inequality \bea \label{e7318} -1
\le \frac{2 s s^\prime + (s + s^\prime + Q^2 )(m_b^2 - m_c^2 - s) +
2 s (m_c^2-m^2_{q})}{\lambda^{1/2} (s,s^\prime,-Q^2)
\lambda^{1/2}(m_b^2,m_c^2,s)} \le +1~, \eea where
$\lambda(a,b,c)=a^2+b^2+c^2-2ab-2ac-2bc$. From this inequality, one
can express $s$ in terms of $s'$ i.e. $f(s')$ in the $s-s'$ plane.

  Straightforward calculations lead to the following expressions  for the
spectral densities:
\begin{eqnarray}\label{11au}
\rho_{V}(s,s',Q^2)&=&N_{c}I_{0}(s,s',-Q^2)\left[-4{m_{c}+4(m_{b}-m_{c})E_{1}+4(m_{q}-m_{c})E_{2}}\right],\nonumber\\
\\
\rho_{A_1}
(s,s',Q^2)&=&N_{c}I_{0}(s,s',-Q^2)[8(m_{c}-m_{b})D_{1}-4m_{b}m_{c}m_{q}\nonumber\\&&+
4(m_{q}+ m_{b}-m_{c })m_{c}^2-2(m_{q}-m_{c})\Delta\nonumber
\\&&-
2(m_{b}-m_{c})\Delta'-2m_{c}u],\nonumber \\
\\
\rho_{A_2}(s,s',Q^2)&=&2N_{c}I_{0}(s,s',-Q^2)[E_2m_b+D_3(m_b-m_c)+(E_1-E_2)m_c\nonumber\\&&+
D_2(m_c-m_b)-E_2m_{q}]~, \nonumber \\
\end{eqnarray}
\begin{eqnarray}\label{110au}
\rho _{V^{\prime }}(s,s^{\prime },Q^{2})
&=&-N_{{c}}I_{0}(s,s^{\prime
},-Q^{2})[4m_{{c}}+4E_{{1}}\left( -m_{{b}}+m_{{c}}\right) +4E_{{2}}\left( m_{%
{c}}+m_{{s}}\right) ],  \nonumber  \nonumber  \\
\\
\rho _{A_{1}^{\prime }}(s,s^{\prime },Q^{2})
&=&-2N_{{c}}I_{0}(s,s^{\prime },-Q^{2})[4\,D_{{1}}\left(
m_{{b}}-m_{{c}}\right) +\Delta ^{\prime }\left(
m_{{b}}-m_{{c}}\right) \nonumber  \\
&&-\Delta \,\left( m_{{c}}+m_{{s}}\right) +2\,{m_{{c}}}^{2}\left( m_{{c}}+m_{%
{s}}-m_{{b}}\right) \nonumber  \\
&&-m_{{c}}\left( 2\,m_{{b}}m_{{s}}-u\right) ],\nonumber  \\
\\
\rho _{A_{2}^{\prime }}(s,s^{\prime },Q^{2})
&=&\,-N_{{c}}I_{0}(s,s^{\prime
},-Q^{2})[2D_{{2}}\left( m_{{b}}-m_{{c}}\right) +2D_{{3}}\left( -m_{{b}}+m_{{%
c}}\right)   \nonumber \\
&&+2E_{{2}}\left( m_{{c}}-m_{{b}}-m_{{s}}\right) -2E_{{1}}m_{{c}}],\nonumber \\
&&
\end{eqnarray}
where  $u=s+s^\prime+Q^2$, $\Delta=s+m_c^2-m_b^2$, $\Delta'=s^\prime
+m_c^2-m^2_{q}$ and $N_c=3$ is the number of colors.  The  functions
$E_1$, $E_2$, $D_1$, $D_2$, $D_3$ and $I_{0}$ are defined as:
\begin{eqnarray}\label{12}
I_{0}(s,s',-Q^2)&=&\frac{1}{4\lambda^{1/2}(s,s',-Q^2)},\nonumber\\
 \lambda(s,s',-Q^2)&=&s^2+s'^2+Q^4+2sQ^2+2s'Q^2-2ss',\nonumber \\
E_{1}&=&\frac{1}{\lambda(s,s',-Q^2)}[2s'\Delta-\Delta'u],\nonumber\\
 E_{2}&=&\frac{1}{\lambda(s,s',-Q^2)}[2s\Delta'-\Delta u],\nonumber\\
 D_{1}&=&\frac{1}{2\lambda(s,s',-Q^2)}[\Delta'^{2}s+\Delta^2s'
 - 4m_{c}^2ss'-\Delta\Delta'u+m_{c}^2u^2],\nonumber\\
 D_{2}&=&\frac{1}{\lambda^{2}(s,s',-Q^2)}[2\Delta'^2ss'+6\Delta^2s'^2
 -8m_{c}^2ss'^2-6\Delta\Delta's'u
 \nonumber \\
 && +\Delta'^2u^2
+2m_{c}^2s'u^2],\nonumber\\
 D_{3}&=&\frac{1}{\lambda^{2}(s,s',-Q^2)}[2\Delta^2ss'+6\Delta'^2s^2
 -8m_{c}^2s's^2-6\Delta\Delta'su
 \nonumber \\
 && +\Delta^2u^2
+2m_{c}^2su^2].\nonumber\\
\end{eqnarray}
\begin{figure}
\vspace*{-1cm}
\begin{center}
\includegraphics[width=10cm]{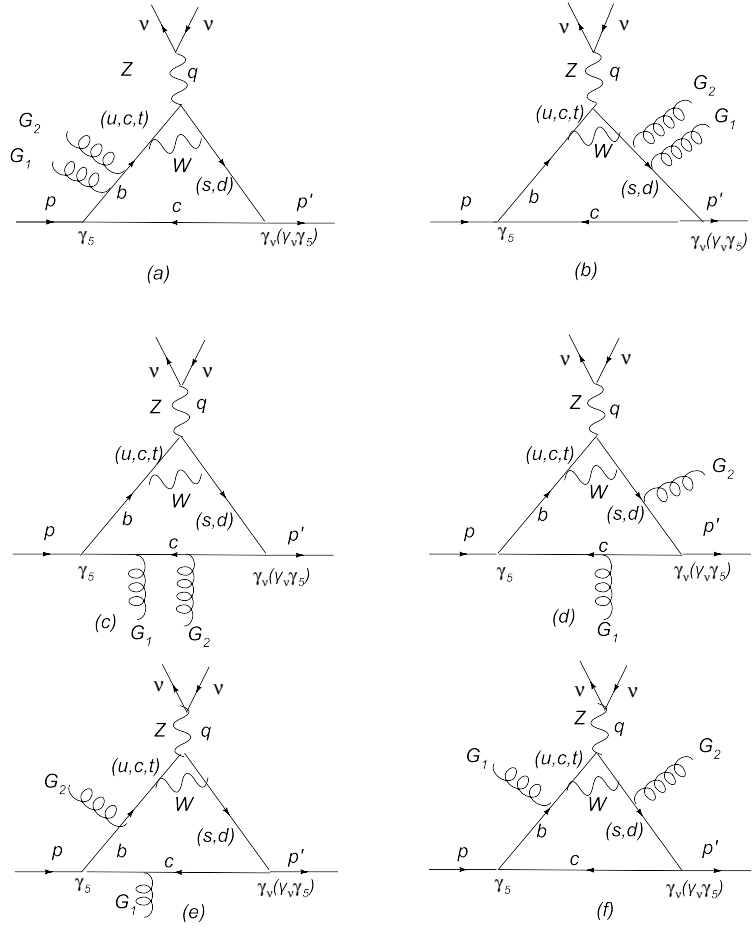}
\end{center}
\caption{Gluon condensate contributions to $B_c \rar X\nu\bar{\nu}$
transitions  } \label{fig2}
\end{figure}

The next step is to calculate  the gluon condensate contributions
to the correlation function (diagrams in Fig. 2). In this section,
we proceed in the definition of the integrals appearing in
evaluation of the gluon condensates contribution the same as in
Refs.~\cite{Aliev3,R7323}. The  diagrams are calculated in the
Fock--Schwinger fixed--point gauge \cite{R7320,R7321,R7322} \bea
x^\mu G_\mu^a = 0~, \eea where $G_\mu^a$ is the gluon field. In
calculating the diagrams,  the following type of integrals are
appeared: \bea \label{e7323} I_0[a,b,c] \es \int \frac{d^4k}{(2
\pi)^4} \frac{1}{\left[ k^2-m_b^2 \right]^a \left[ (p+k)^2-m_c^2
\right]^b \left[ (p^\prime+k)^2-m_{q}^2\right]^c}~,
\nnb \\ \nnb \\
I_\mu[a,b,c] \es \int \frac{d^4k}{(2 \pi)^4} \frac{k_\mu}{\left[
k^2-m_b^2 \right]^a \left[ (p+k)^2-m_c^2 \right]^b \left[
(p^\prime+k)^2-m_{q}^2\right]^c}~,
\nnb \\ \nnb \\
I_{\mu\nu}[a,b,c] \es \int \frac{d^4k}{(2 \pi)^4} \frac{k_\mu
k_\nu}{\left[ k^2-m_b^2 \right]^a \left[ (p+k)^2-m_c^2 \right]^b
\left[ (p^\prime+k)^2-m_{q}^2\right]^c}~. \eea

Hear, $k$ is the momentum of the spectator quark $c$. In Schwinger
representation for the propagators, i.e., \bea \label{e7324}
\frac{1}{p^2+m^2} = \frac{1}{\Gamma(\alpha)} \int_0^\infty d\alpha
\, \alpha^{n-1} e^{-\alpha(p^2+m^2)}~, \eea the integrals take the
suitable form to apply   the Borel transformations as  \bea
\label{e7325} {\cal B}_{\hat{p}^2} (M^2) e^{-\alpha p^2} = \delta
(1/M^2-\alpha)~. \eea Performing all integrals and applying double
Borel transformations with respect to $p^2$ and $p^{\prime 2}$ the
Borel transformed form of the integrals are obtained as \bea
\label{e7326} \hat{I}_0(a,b,c) \es \frac{(-1)^{a+b+c}}{16
\pi^2\,\Gamma(a) \Gamma(b) \Gamma(c)}
(M_1^2)^{2-a-b} (M_2^2)^{2-a-c} \, {\cal U}_0(a+b+c-4,1-c-b)~, \nnb \\ \nnb \\
\hat{I}_\mu(a,b,c) \es \frac{1}{2} \Big[\hat{I}_1(a,b,c) +
\hat{I}_2(a,b,c)\Big] {\cal P}_\mu + \frac{1}{2} \Big[\hat{I}_1(a,b,c) -
\hat{I}_2(a,b,c)\Big] q_\mu~, \nnb \\ \nnb \\
\hat{I}_{\mu\nu}(a,b,c) \es \hat{I}_6(a,b,c) g_{\mu\nu} +
\frac{1}{4} \Big(2 \hat{I}_4 + \hat{I}_3 + \hat{I}_5 \Big) {\cal P}_\mu {\cal P}_\nu
+ \frac{1}{4} \Big(-\hat{I}_5 + \hat{I}_3 \Big) {\cal P}_\mu q_\nu \nnb \\
\ar \frac{1}{4} \Big(-\hat{I}_5 + \hat{I}_3 \Big) {\cal P}_\nu q_\mu +
\frac{1}{4} \Big(-2 \hat{I}_4 + \hat{I}_3 + \hat{I}_5 \Big) q_\mu q_\nu~,
\eea
where
\bea
\label{e7327}
\hat{I}_1(a,b,c) \es i \frac{(-1)^{a+b+c+1}}{16 \pi^2\,\Gamma(a) \Gamma(b) \Gamma(c)}
(M_1^2)^{2-a-b} (M_2^2)^{3-a-c} \, {\cal U}_0(a+b+c-5,1-c-b)~, \nnb \\ \nnb \\
\hat{I}_2(a,b,c) \es i \frac{(-1)^{a+b+c+1}}{16 \pi^2\,\Gamma(a) \Gamma(b) \Gamma(c)}
(M_1^2)^{3-a-b} (M_2^2)^{2-a-c} \, {\cal U}_0(a+b+c-5,1-c-b)~, \nnb \\ \nnb \\
\hat{I}_3(a,b,c) \es i \frac{(-1)^{a+b+c}}{32 \pi^2\,\Gamma(a) \Gamma(b) \Gamma(c)}
(M_1^2)^{2-a-b} (M_2^2)^{4-a-c} \, {\cal U}_0(a+b+c-6,1-c-b)~,\nnb \\ \nnb \\
\hat{I}_4(a,b,c) \es i \frac{(-1)^{a+b+c}}{16 \pi^2\,\Gamma(a) \Gamma(b) \Gamma(c)}
(M_1^2)^{3-a-b} (M_2^2)^{3-a-c} \, {\cal U}_0(a+b+c-6,1-c-b)~,\nnb \\ \nnb \\
\hat{I}_5(a,b,c) \es i \frac{(-1)^{a+b+c}}{16 \pi^2\,\Gamma(a) \Gamma(b) \Gamma(c)}
(M_1^2)^{4-a-b} (M_2^2)^{2-a-c} \, {\cal U}_0(a+b+c-6,1-c-b)~,\nnb \\ \nnb \\
\hat{I}_6(a,b,c) \es i \frac{(-1)^{a+b+c+1}}{16 \pi^2\,\Gamma(a)
\Gamma(b) \Gamma(c)} (M_1^2)^{3-a-b} (M_2^2)^{3-a-c} \, {\cal
U}_0(a+b+c-6,2-c-b)~. \eea Here, hat in Eqs. (\ref{e7326}) and
(\ref{e7327}) denotes  the double Borel transformed form of
integrals. $M_1^2$ and $M_2^2$ are the Borel parameters in the $s$
and $s^\prime$ channels, respectively, and the function ${\cal
U}_0(\alpha,\beta)$ is defined in the following way \bea {\cal
U}_0(a,b) = \int_0^\infty dy (y+M_1^2+M_2^2)^a y^b \,exp\left[
-\frac{B_{-1}}{y} - B_0 - B_1 y \right]~, \nnb \eea where \bea
\label{e7328} B_{-1} \es \frac{1}{M_1^2M_2^2}
\left[m_{q}^2M_1^4+m_b^2 M_2^4 + M_2^2M_1^2 (m_b^2+m_{q}^2
+ Q^2) \right] ~, \nnb \\
B_0 \es \frac{1}{M_1^2 M_2^2} \left[ (m_{q}^2+m_c^2) M_1^2 + M_2^2
(m_b^2+m_c^2)
\right] ~, \nnb \\
B_{1} \es \frac{m_c^2}{M_1^2 M_2^2}~. \eea After lengthy
calculations, the following results for the gluon condensate
contributions  are obtained:
\begin{equation}
\Pi_{i(i')}^{\la G^2 \ra}=-i \lla \frac{\alpha_s}{\pi} G^2 \rra
\frac{C_{i(i')}}{12},
\end{equation}
where the explicit expressions for $C_{i(i')}$ and  are given in
appendix--A.

 Applying  double Borel
transformations with respect to the $p^2$ ($p^2\rightarrow
M_{1}^{2}$) and $p'^2$ ($p^{\prime 2}\rightarrow M_{2}^{2}$) on the
phenomenological as well as the perturbative  parts of the
correlation  function  and equating the physical and QCD sides of
the correlator, the following sum rules for the form factors
 $V$, $A_1$ and $A_2$ are obtained: \bea
V (V')\es  -\frac{(m_b+m_c) (m_{B_c} + m_{X})}{8\pi^2 f_{B_c}
m_{B_c}^2 f_{X} m _{X}} e^{m_{B_c}^2/M_1^2}
e^{m_{X}^2/M_2^2} \nnb \\
\cp  \Bigg\{ \int_{(m_c+m_{q})^2}^{s_0^\prime} ds^\prime
\int_{f_{-}(s')}^{min(s_0,f_{+}(s'))}ds \rho_{V(V')}
(s,s^\prime,Q^2) e^{-s/M_1^2} e^{-s^\prime/M_2^2} - i  \lla
\frac{\alpha_s}{\pi} G^2
\rra \frac{C_{V(V')}}{12} \Bigg\}~, \nnb \\ \nnb \\
A_1(A'_1) \es -\frac{(m_b+m_c)}{4\pi^2f_{B_c} m_{B_c}^2 f_{X} m
_{X}(m_{B_c} + m_{X})} e^{m_{B_c}^2/M_1^2}
e^{m_{X}^2/M_2^2} \nnb \\
\cp  \Bigg\{ \int_{(m_c+m_{q})^2}^{s_0^\prime} ds^\prime
\int_{f_{-}(s')}^{min(s_0,f_{+}(s'))} ds\rho_{A_1(A'_1)}
(s,s^\prime,Q^2) e^{-s/M_1^2} e^{-s^\prime/M_2^2} - i \lla
\frac{\alpha_s}{\pi} G^2
\rra \frac{C_{A_1(A'_1)}}{12} \Bigg\}~, \nnb \\ \nnb \\
A_2(A'_2) \es \frac{m_{X}(m_b+m_c)(m_{B_c} + m_{X})}{\pi^2 f_{B_c}
m_{B_c}^2 f_{X} (m_{B_c}^2+3m _{X}^{2}+Q^{2})} e^{m_{B_c}^2/M_1^2}
e^{m_{X}^2/M_2^2} \nnb \\
\cp  \Bigg\{\int_{(m_c+m_{q})^2}^{s_0^\prime} ds^\prime
\int_{f_{-}(s')}^{min(s_0,f_{+}(s'))}ds \rho_{A_2(A'_2)}
(s,s^\prime,Q^2) e^{-s/M_1^2} e^{-s^\prime/M_2^2} - i\lla
\frac{\alpha_s}{\pi} G^2 \rra \frac{C_{A_2(A'_2)}}{12} \Bigg\}~,\nnb \\
\eea
 where $s_0$ and $s_0^\prime$ are the continuum thresholds in
 $B_c$ and $X$ channels, respectively  and $f_{\pm}(s')$
in the lower and upper limit of the integral  over $s$ are obtained
from inequality ( \ref{e7318}) with respect to $s$, i.e.,
$s=f_{\pm}(s')$. By $min(s_0,f_{+}(s'))$, for each value of the
$q^{2}$, between $s_{0}$ and $f_{+}$, the smaller one is selected.
In above equation, in order to subtract the contributions of the
higher states and the continuum the quark-hadron duality assumption
is also
 used, i.e., it is assumed
that
\begin{eqnarray}
\rho^{higher states}(s,s') = \rho^{OPE}(s,s') \theta(s-s_0)
\theta(s-s'_0).
\end{eqnarray}
In physical side and perturbative part of the correlation function,
we also use the following Borel transformations
\begin{eqnarray}
{\cal B}_{p^{2}}\left\{\frac{1}{m^2(s)-p^2}\right\}& =&
e^\frac{m^2(s)}{M_{1}^2},\nonumber
\\
{\cal B}_{p'^2}\left\{\frac{1}{m^2(s')-p'^2}\right\}& =&
e^\frac{m^2(s')}{M_{2}^2}.
\end{eqnarray}
 At the end of this section, we would like to present the
differential decay width of  $B_c \rar X\nu\bar{\nu}$ decays in
terms of the form factors. Using the amplitude in Eq. (\ref{2au}),
we obtain the following expressions for the differential decay width
of these transitions.
\begin{eqnarray}\label{reza1}
 \frac{d{\Gamma}}{dq^2}(B_{c}
\rightarrow V {\nu} \bar \nu)\nonumber &=&
\frac{G_{_F}^2{\alpha}^2}{2^{10}{\pi}^5}{\mid}V_{tq}V^*_{tb}{\mid}^2
{\lambda}^{1/2}(1,r_{_V},t')m_{B_{c}}^3{\mid}C_{10}{\mid}^2 \\
&\times& \Bigg( 8{\lambda(1,r_{_V},t')}t'
\frac{V^2}{(1+\sqrt{r_{_V}})^2}+\frac{1}{r_{_V}}\bigg[{\lambda(1,r_{_V},t')}^2
\frac{A_2^2}{(1+\sqrt{r_{_V}})^2} \nonumber \\
&+& (1+\sqrt{r_{_V}})^2({\lambda(1,r_{_V},t')}+12 r_{_V} t')A_1^2
-2{\lambda(1,r_{_V},t')}(1- r_{_V} - t') Re(A_1A_2) \bigg]
\Bigg),\nonumber\\
\end{eqnarray}
\begin{eqnarray}\label{reza2}
 \frac{d{\Gamma}}{dq^2}(B_{c}
\rightarrow AV {\nu} \bar \nu)\nonumber &=&
\frac{G_{_F}^2{\alpha}^2}{2^{10}{\pi}^5}{\mid}V_{tq}V^*_{tb}{\mid}^2
{\lambda}^{1/2}(1,r_{V'},t')m_{B_{c}}^3{\mid}C_{10}{\mid}^2 \\
&\times& \Bigg( 8{\lambda(1,r_{V'},t')}t'
\frac{V^{'2}}{(1+\sqrt{r_{V'}})^2}+\frac{1}{r_{V'}}\bigg[{\lambda(1,r_{V'},t')}^2
\frac{A_2^{'2}}{(1+\sqrt{r_{V'}})^2} \nonumber \\
&+& (1+\sqrt{r_{_V'}})^2({\lambda(1,r_{V'},t')}+12 r_{_V'}
t')A_1^{'2} -2{\lambda(1,r_{V'},t')}(1- r_{V'} - t') Re(A'_1A'_2)
\bigg]
\Bigg),\nonumber\\
\end{eqnarray}
where, ${\lambda}(1,r_{V},t')$ and ${\lambda}(1,r_{V'},t')$  are the
usual triangle function with
\begin{equation}
{\lambda}(a,b,c)=a^2+b^2 +c^2-2ab-2ac-2bc ,\quad ~~\end{equation}
and
\begin{equation}
~~~~ r_{_V}=\frac{m_{_V}^2}{m_{B_{c}}^2},\quad ~~~~
r_{_V'}=\frac{m_{AV}^2}{m_{B_{c}}^2},\quad
t'=-\frac{Q^2}{m_{B_{c}}^2} .
\end{equation}
The total decay widths  are obtained from  integration  of Eqs.
(\ref{reza1}) and (\ref{reza2})on $q^2$ in the interval
$0<q^2<(m_{B_{c}}-m_{V(AV)})^2$.

\section{Numerical analysis}
The explicit expressions for the form factors  $V$, $A_{1}$,
$A_{2}$, $V'$, $A'_{1}$ and  $A'_{2}$ and
$\frac{d{\Gamma}}{dq^2}(B_{c} \rightarrow X {\nu} \bar
\nu)\nonumber$ indicate that the main input parameters entering to
the expressions are Gluon condensate, Wilson coefficient $C_{10}$ ,
elements of the CKM matrix $V_{tb}$, $V_{ts}$ and $V_{td}$, leptonic
decay constants; $f_{B_{C}}$, $f_{D^{\ast}}$, $f_{D_{s}^{\ast}}$ and
$f_{D_{s1}}$, Borel parameters $M_{1}^2$ and $M_{2}^2$, as well as
the continuum thresholds $s_{0}$ and $s'_{0}$. For the numerical
values of   the Gluon condensate, leptonic decay constants, CKM
matrix elements, Wilson coefficient and quark and meson masses
$<\frac{\alpha_{s}}{\pi}G^{2}>=0.012~ GeV ^{4}$ \cite{Shifman1},
  $C_{10}=-4.669$
 \cite{Buras,Bashiry},  $\mid
V_{tb}\mid=0.77^{+0.18}_{-0.24}$, $\mid
V_{ts}\mid=(40.6\pm2.7)\times10^{-3}$ $\mid
V_{td}\mid=(7.4\pm0.8)\times10^{-3}$ \cite {Ceccucci},
$f_{D_{s}^{\ast}} =266\pm32
  ~MeV $ \cite{Colangelo1},
  $f_{D^{\ast}}=0.23\pm0.02~GeV$,
\cite{Bowler}, $f_{B_{c}} =350
  ~MeV $ \cite{Colangelo2,Kiselev,Alievy},  $f_{D_{s1}}=0.225\pm0.025~GeV$ \cite{Colangelo1}, $ m_{c}(\mu=m_{c})=
 1.275\pm
 0.015~ GeV$, $m_{s}(1~ GeV)\simeq 142 ~MeV$ \cite{Huang} ,  $m_{b} =
(4.7\pm
 0.1)~GeV$ \cite{Ioffe}, $m_{d}=(3-7)~MeV$, $m_{D_{s}^{\ast}}=2.112~GeV$,
 $m_{D^{\ast}}=2.010~GeV$,  $m_{D_{s1}}=2.460~GeV$
  and $ m_{B_{C}}=6.258~GeV$ \cite{Yao} are used.

The expressions for the form factors  contain also four auxiliary
parameters: Borel mass squares $M_{1}^2$ and $M_{2}^2$ and continuum
threshold $s_{0}$  and $s'_{0}$. These are not physical quantities,
hence the physical quantities, form factors, must be independent of
these auxiliary parameters. We should find the "working regions" of
these  parameters, where the form factors are independent of them.
The parameters $s_0$ and $s_0^\prime$, which are the continuum
thresholds of $B_c$ and $X$ mesons, respectively, are
 determined from the conditions that
guarantees the sum rules to have the best stability in the allowed
$M_1^2$ and $M_2^2$ region. The values of continuum thresholds
calculated from the two--point QCD sum rules are taken to be
$s_0=45~GeV^2$ and $s_0^\prime=8~GeV^2$
\cite{Aliev1,Shifman1,Colangelo1}.  The working regions for $M_1^2$
and $M_2^2$ are determined by requiring that from one side, the
continuum and higher states contributions are effectively suppressed
and other side the gluon condensate contribution is small, which
guarantees that the contributions of higher dimensional operators
are small. Both conditions are satisfied in the  regions $10~GeV^2
\le M_1^2 \le 25~GeV^2$ and $5~GeV^2 \le M_2^2 \le 15~GeV^2$.

The values of  the form factors at $q^2=0$ are  shown in Table 1:
\begin{table}[h]
\centering
\begin{tabular}{|c|c|c|c|
} \hline
  &$ B_c \rightarrow D^* \nu\bar{\nu}$  & $B_c \rightarrow D^*_s \nu\bar{\nu}$&$ B_c \rightarrow D_{s1}(2460)\nu\bar{\nu}$   \\ \cline{1-4}
 $V(0)$ &$0.27\pm0.016$ & $0.29\pm0.017$ &$0.30\pm0.017$ \\\cline{1-4}
 $A_{1}(0)$ &$0.12\pm0.012$ & $0.15\pm0.014 $& $0.13\pm0.012$ \\\cline{1-4}
 $A_{2}(0)$ &$-0.018\pm0.0016$ & $-0.03\pm0.002$&$-0.07\pm0.005$   \\\cline{1-4}

 \end{tabular}
 \vspace{0.8cm}
\caption{The values of  the form factors at $q^2=0$, for
$M_{1}^2=18~GeV^2$, $M_{2}^2=8~GeV^2$ .}\label{tab:1}
\end{table}

For obtaining  the $q^2$ dependent expressions of the form factors ,
we should consider a range of $ q^2$ where the sum rules can
reliably be calculated. Our sum rules for the form factors are
truncated at  $(1.2-2)~GeV$ below  the perturbative cut. In order to
extend our results to the full physical region, i.e., the regions $0
\leq q^2 \leq 18~ GeV^2$, $0 \leq q^2 \leq 17.2~ GeV^2$ and $0 \leq
q^2 \leq 14.4~ GeV^2$ for $ B_c \rightarrow D^* \nu\bar{\nu}$, $B_c
\rightarrow D^*_s \nu\bar{\nu}$ and $B_c \rightarrow D_{s1}
\nu\bar{\nu}$, respectively,  we look for parameterization of the
form factors in such a way that
 this parameterization coincides
with the sum rules prediction. Our numerical calculations shows that
the best parameterization of the
form factors with respect to $-Q^2$ are as follows:\\
 \begin{equation}\label{17au}
 f_{i}(-Q^2)=\frac{f_{i}(0)}{1+ \alpha\hat{q}+ \beta\hat{q}^2},
\end{equation}
where $\hat{q}=-Q^2/m_{B_{c}}^2$. The values of the parameters
 $f_{i}(0)$, $\alpha$ and $\beta$ are
given in the Tables 2,  3 and 4 for $ B_c \rightarrow D^*
\nu\bar{\nu}$, $B_c \rightarrow D^*_s \nu\bar{\nu}$ and $B_c
\rightarrow D_{s1} \nu\bar{\nu}$, respectively.
\begin{table}[h]
\centering
\begin{tabular}{|c|c|c|c|} \hline
  & f(0)  & $ \alpha$ & $ \beta$\\\cline{1-4}
 $V$ & 0.27 &-7.76   & 22.83 \\\cline{1-4}
 $A_{1}$ & 0.12  & -8.37 & 22.29 \\\cline{1-4}
 $A_{2}$ & -0.018  & -11.63 & 33.52 \\\cline{1-4}

 \end{tabular}
 \vspace{0.8cm}
\caption{Parameters appearing in the form factors of the
$B_{c}\rightarrow D^* \nu \bar{\nu}$}decay in a two-parameter fit,
for $M_{1}^2=18~GeV^2$, $M_{2}^2=8~GeV^2.$ \label{tab:2}
\end{table}
\begin{table}[h]
\centering
\begin{tabular}{|c|c|c|c|} \hline
  & f(0)  & $ \alpha$ & $ \beta$\\\cline{1-4}
 $V$ & 0.29 &-3.17  & 9.90\\\cline{1-4}
 $A_{1}$ & 0.15  & -3.60 & 8.13\\\cline{1-4}
 $A_{2}$ & -0.03  & -3.29 & 10.85\\\cline{1-4}

 \end{tabular}
 \vspace{0.8cm}
\caption{Parameters appearing in the form factors of the
$B_{c}\rightarrow D^*_s\nu \bar{\nu}$}decay in a two-parameter fit,
for $M_{1}^2=18~GeV^2$, $M_{2}^2=8~GeV^2.$ \label{tab:3}
\end{table}
\begin{table}[h]
\centering
\begin{tabular}{|c|c|c|c|} \hline
  & f(0)  & $ \alpha$ & $ \beta$\\\cline{1-4}
 $V$ &0.30 &-1.30   &4.35  \\\cline{1-4}
 $A_{1}$ &0.13   &-2.40  &4.43  \\\cline{1-4}
 $A_{2}$ &-0.07   &-1.25  & 4.30 \\\cline{1-4}

 \end{tabular}
 \vspace{0.8cm}
\caption{Parameters appearing in the form factors of the
$B_{c}\rightarrow D_{s1} \nu \bar{\nu}$}decay in a two-parameter
fit, for $M_{1}^2=18~GeV^2$, $M_{2}^2=8~GeV^2.$ \label{tab:4}
\end{table}

At the end of this section, we would like to present the value of
the branching ratio of these decays.
 Taking into account the $q^2$ dependencies of
the form factors and performing integration over $q^2$  in Eqs.
(\ref{reza1}), (\ref{reza2})  in the whole physical region and using
the total life time of $B_c$ meson, $\tau \simeq 0.46~ps$
\cite{R7326}, the branching ratio of the $B_c \rar X\nu\bar{\nu}$
decays are obtained as presented in Table 5. This Table also
encompasses a comparison of our results with the existing
predictions of the relativistic constituent quark
 model (RCQM)
\cite{tt1}:
\begin{table}[h]
\centering
\begin{tabular}{|c|c|c|
} \hline
 decay & our result  & RCQM
\cite{tt1}\\ \cline{1-3}
 $Br(B_c \rightarrow D^* \nu\bar{\nu})*10^{-8}$ &$5.23\pm0.12$ & 5.78  \\\cline{1-3}
 $Br(B_c \rightarrow D^*_s \nu\bar{\nu})*10^{-6}$ &$1.34\pm0.25$ &1.42  \\\cline{1-3}
$Br(B_c \rightarrow D_{s1} \nu\bar{\nu})*10^{-6}$ &1.73$\pm0.10$ &-
\\\cline{1-3}
 \end{tabular}
 \vspace{0.8cm}
\caption{Our results for the branching ratios and their comparisons
with the prediction of the relativistic constituent quark
 model (RCQM)
\cite{tt1}}. \label{tab:5}
\end{table}

 From this Table, we see  a good consistency between
our results and that of the relativistic constituent quark
 model.

In summary, we investigated the rare $B_c \rar X \nu\bar{\nu}$
transition with $X$ been axial vector particle, $AV(D_{s1})$, and
vector particles, $V(D^*,D^*_s)$  in the framework of the three
point QCD sum rules. The $q^2$ dependent expressions for the form
factors were calculated. The quark condensates contributions to the
correlation function were zero, so we considered the gluon
corrections to the correlation function as a first nonperturbative
contributions. Finally, we calculated the total decay width and
branching ratio of these decays and compared our results with the
predictions of the quark model. Our results are in good agreement
with the relativistic constituent quark model.
\section*{Acknowledgments}
The authors would like to thank T. M. Aliev and A. Ozpineci for
their useful discussions. One of the authors (K. Azizi) thanks
Turkish Scientific and Research Council (TUBITAK) for their partial
financial support.

\newpage

\appendix
\begin{center}
{\Large{\bf Appendix--A}}
\end{center}


\setcounter{equation}{0}
\renewcommand{\theequation}{C.\arabic{equation}}
\setcounter{section}{0} \setcounter{table}{0}

\section*{}
In this appendix we give the explicit expressions of the
coefficients of the gluon condensate which enter to the sum rules
for the form factors $V$, $A_1$ and $A_2$, respectively.
\begin{eqnarray*}
&&C_{V}=-20\,I_{{1}}(3,2,2){m_c}^{5}-20\,I_{{2}}(3,2,2){\mathit{%
m_c}}^{5}-20\,I_{{0}}(3,2,2){m_c}^{5}+20\,I_{{1}}(3,2,2){m_c}%
^{4}m_b \\
&&+20\,I_{{1}}(3,2,2){m_c}^{3}{m_b}^{2}+20\,I_{{0}}(3,2,2){%
m_c}^{3}{m_b}^{2}+20\,I_{{2}}(3,2,2){m_c}^{3}{%
m_b}^{2}-20\,I_{{1}}(3,2,2){m_c}^{2}{m_b}^{3} \\
&&-40\,I_{{0}}(2,2,2){m_c}^{3}-20\,I_{{2}}(3,1,2){m_c}%
^{3}+40\,I_{1}^{[0,1]}(3,2,2){m_c}^{3}-20\,I_{{0}}(3,1,2){m_c%
}^{3} \\
&&+40\,I_{2}^{[0,1]}(3,2,2){m_c}^{3}-60\,I_{{2}}(4,1,1){m_c}%
^{3}-40\,I_{{1}}(2,2,2){m_c}^{3}+20\,I_{{2}}(3,2,1){m_c}^{3}
\\
&&-40\,I_{{2}}(2,2,2){m_c}^{3}-60\,I_{{0}}(4,1,1){m_c}%
^{3}+40\,I_{0}^{[0,1]}(3,2,2){m_c}^{3}-60\,I_{{1}}(4,1,1){m_c%
}^{3} \\
&&-20\,I_{{2}}(3,2,1){m_c}^{2}m_b+60\,I_{{1}}(4,1,1){\mathit{%
m_c}}^{2}m_b+80\,I_{{0}}(2,3,1){m_c}^{2}m_b+20\,I_{{0}%
}(3,2,1){m_c}^{2}m_b \\
&&+40I_{{1}}(2,2,2){m_c}^{2}m_b+20\,I_{{1}}(3,2,1){\mathit{mc%
}}^{2}m_b-40\,I_{1}^{[0,1]}(3,2,2){m_c}^{2}m_b%
+40\,I_{{1}}(2,3,1){m_c}^{2}m_b \\
&&+120\,I_{{0}}(1,4,1)m_c\,{m_b}^{2}+20\,I_{2}^{[0,1]}(3,2,2)%
m_c\,{m_b}^{2}+20\,I_{0}^{[0,1]}(3,2,2)m_c\,{\mathit{%
m_b}}^{2}+20\,I_{1}^{[0,1]}(3,2,2)m_c\,{m_b}^{2} \\
&&+120\,I_{{1}}(1,4,1)m_c\,{m_b}^{2}-40\,I_{{0}}(3,2,1)%
m_c\,{m_b}^{2}-40\,I_{{2}}(3,2,1)m_c\,{m_b}%
^{2}-60\,I_{{1}}(3,2,1)m_c\,{m_b}^{2} \\
&&-120\,I_{{1}}(1,4,1){m_b}^{3}-40\,I_{{1}}(2,3,1){m_b}%
^{3}+40\,I_{{1}}(3,2,1){m_b}^{3}+20\,I_{{1}}(2,2,2){m_b}^{3}
\\
&&-20\,I_{1}^{[0,1]}(3,2,2){m_b}^{3}+40\,I_{1}^{[0,1]}(3,1,2)\mathit{%
m_c}+40\,I_{1}^{[0,1]}(3,2,1)m_c+40\,\,I_{2}^{[0,1]}(2,2,2)m_c
\\
&&+60\,I_{2}^{[0,1]}(3,1,2)m_c-20\,I_{2}^{[0,2]}(3,2,2)m_c%
-20\,I_{{1}}(3,1,1)m_c-20\,I_{1}^{[0,2]}(3,2,2)m_c \\
&&+40\,I_{1}^{[0,1]}(2,2,2)m_c+40I_{1}^{[0,1]}(3,2,1)m_c%
-20\,I_{{0}}(3,1,1)m_c+20\,I_{{2}}(3,1,1)m_c \\
&&-60\,I_{{0}}(2,1,2)m_c-40\,I_{{1}}(1,2,2)m_c%
+20\,I_{2}^{[0,1]}(3,2,1)m_c-20\,I_{0}^{[0,2]}(3,2,2)m_c \\
&&-60\,I_{{2}}(2,1,2)m_c+60\,I_{0}^{[0,1]}(3,1,2)m_c-40\,I_{{%
2}}(1,2,2)m_c+40\,I_{{0}}(2,2,1)m_c \\
&&+60\,I_{{2}}(2,2,1)m_c-40\,I_{{1}}(2,1,2)m_c-40\,I_{{0}%
}(1,2,2)m_c+40\,I_{{1}}(2,2,1)m_c \\
&&+40\,I_{0}^{[0,1]}(2,2,2)m_c-40\,I_{{0}}(2,2,1)m_b%
-40\,I_{1}^{[0,1]}(3,1,2)m_b+80\,I_{2}^{[0,1]}(2,3,1)m_b \\
&&-100\,I_{{1}}(2,2,1)m_b+60\,I_{{1}}(2,1,2)m_b%
-40\,I_{1}^{[0,1]}(3,2,1)m_b+40\,I_{{1}}(1,2,2)m_b \\
&&+200\,I_{{0}}(1,3,1)m_b+40\,I_{1}^{[0,1]}(2,3,1)m_b%
+120\,I_{{2}}(1,3,1)m_b+20\,I_{1}^{[0,2]}(3,2,2)m_b \\
&&-40\,I_{{2}}(2,2,1)m_b-40\,I_{1}^{[0,1]}(2,2,2)m_b+40\,I_{{%
1}}(1,3,1)m_b   ~,
\end{eqnarray*}
\begin{eqnarray*}
&&C_{A_{1}}=10\,I_{{0}}(3,2,2){m_c}^{6}{m_b}-10\,I_{{0}%
}(3,2,2){m_c}^{5}{m_b}^{2}-10\,I_{{0}}(3,2,2){m_c}%
^{4}{m_b}^{3}+10\,I_{{0}}(3,2,2){m_c}^{3}{m_b}^{4} \\
&&+40\,I_{{6}}(3,2,2){m_c}^{5}+30\,I_{{0}}(2,2,2){m_c}^{4}%
m_b+20\,I_{{0}}(2,3,1){m_c}^{4}m_b-40\,I_{{6}}(3,2,2)%
{m_c}^{4}m_b \\
&&-30\,I_{0}^{[0,1]}(3,2,2){m_c}^{4}m_b+10\,I_{{0}}(3,2,1){%
m_c}^{4}m_b+30\,I_{{0}}(4,1,1){m_c}^{4}m_b%
-40\,I_{{6}}(3,2,2){m_c}^{3}{m_b}^{2} \\
&&+20\,I_{0}^{[0,1]}(3,2,2){m_c}^{3}{m_b}^{2}-10\,I_{{0}%
}(3,2,1){m_c}^{3}{m_b}^{2}-30\,I_{{0}}(4,1,1){m_c}%
^{3}{m_b}^{2}-20\,I_{{0}}(2,2,2){m_c}^{3}{m_b}^{2} \\
&&-60\,I_{{0}}(1,4,1){m_c}^{2}{m_b}^{3}+40\,I_{{6}}(3,2,2){%
m_c}^{2}{m_b}^{3}+20\,I_{{0}}(3,2,1){m_c}^{2}{%
m_b}^{3}-20\,I_{{0}}(2,3,1){m_c}^{2}{m_b}^{3} \\
&&-20\,I_{{0}}(3,2,1)m_c\,{m_b}^{4}+10\,I_{0}^{[0,1]}(3,2,2)%
m_c\,{m_b}^{4}+60\,I_{{0}}(1,4,1)m_c\,{m_b}%
^{4}+80\,I_{{6}}(3,2,1){m_c}^{3} \\
&&-10\,I_{{0}}(3,1,1){m_c}^{3}+40\,I_{{6}}(3,1,2){m_c}%
^{3}+120\,I_{{6}}(4,1,1){m_c}^{3}-80\,I_{6}^{[0,1]}(3,2,2){\mathit{mc%
}}^{3} \\
&&+80\,I_{{6}}(2,2,2){m_c}^{3}+30\,I_{{0}}(2,1,2){m_c}^{2}%
m_b+20\,I_{{0}}(3,1,1){m_c}^{2}m_b-30\,I_{{0}}(2,2,1)%
{m_c}^{2}m_b \\
&&-20\,I_{0}^{[0,1]}(3,1,2){m_c}^{2}m_b-60%
\,I_{0}^{[0,1]}(2,2,2){m_c}^{2}m_b-40\,I_{0}^{[0,1]}(3,2,1){%
m_c}^{2}m_b-120\,I_{{6}}(4,1,1){m_c}^{2}m_b
\\
&&-30\,I_{0}^{[0,1]}(4,1,1){m_c}^{2}m_b-80\,I_{{6}}(3,2,1){%
m_c}^{2}m_b-40\,I_{0}^{[0,1]}(2,3,1){m_c}^{2}\mathit{%
m_b}+80\,I_{6}^{[0,1]}(3,2,2){m_c}^{2}m_b \\
&&-40\,I_{{6}}(3,1,2){m_c}^{2}m_b+40\,I_{{0}}(1,2,2){\mathit{%
m_c}}^{2}m_b-20\,I_{{0}}(1,3,1){m_c}^{2}m_b-40\,I_{{6}%
}(2,2,2){m_c}^{2}m_b \\
&&+30\,I_{0}^{[o,2]}(3,2,2){m_c}^{2}m_b+80\,I_{{6}}(2,3,1){%
m_c}^{2}m_b-10\,I_{{0}}(3,1,1)m_c\,{m_b}%
^{2}+40\,I_{{6}}(2,2,2)m_c\,{m_b}^{2} \\
&&-30\,I_{{0}}(2,1,2)m_c\,{m_b}^{2}-40\,I_{6}^{[0,1]}(3,2,2)%
m_c\,{m_b}^{2}-240\,I_{{6}}(1,4,1)m_c\,{m_b}%
^{2}+20\,I_{{0}}(2,2,1)m_c\,{m_b}^{2} \\
&&+60\,I_{{0}}(1,3,1)m_c\,{m_b}^{2}+10\,I_{0}^{[0,1]}(3,2,1)%
m_c\,{m_b}^{2}+40\,I_{{6}}(3,2,1)m_c\,{m_b}%
^{2}+20\,I_{0}^{[0,1]}(2,2,2)m_c\,{m_b}^{2} \\
&&-20\,I_{{0}}(1,2,2)m_c\,{m_b}^{2}+40\,I_{0}^{[0,1]}(3,1,2)%
m_c\,{m_b}^{2}-10\,I_{0}^{[0,2]}(3,2,2)m_c\,{\mathit{%
m_b}}^{2}-30\,I_{0}^{[0,1]}(3,2,1){m_b}^{3} \\
&&+40\,I_{{0}}(1,3,1){m_b}^{3}-80\,I_{{6}}(2,3,1){m_b}%
^{3}+10\,I_{0}^{[0,2]}(3,2,2){m_b}^{3}-10\,I_{{0}}(2,2,1){m_b%
}^{3} \\
&&+40\,I_{6}^{[0,1]}(3,2,2){m_b}^{3}+10\,I_{{0}}(1,2,2){m_b}%
^{3}+240\,I_{{6}}(1,4,1){m_b}^{3}-20\,I_{0}^{[0,1]}(2,2,2){\mathit{mb%
}}^{3} \\
&&-40\,I_{{6}}(2,2,2){m_b}^{3}-40\,I_{{6}}(3,2,1){m_b}%
^{3}+60\,I_{0}^{[0,1]}(1,4,1){m_b}^{3}-20\,I_{0}^{[0,1]}(2,3,1){%
m_b}^{3} \\
&&-80\,I_{6}^{[0,1]}(2,2,2)m_c-40\,I_{{6}}(2,1,2)m_c%
+40\,I_{6}^{[0,1]}(3,2,2)m_c+10\,I_{{0}}(2,1,1)m_c \\
&&-10\,I_{{0}}(1,1,2)m_c-10\,I_{0}^{[0,1]}(3,1,1)m_c-40\,I_{{%
6}}(2,2,1)m_c-80\,I_{6}^{[0,1]}(3,2,1)m_c \\
&&-80\,I_{{6}}(3,1,1)m_c-10\,I_{{0}}(1,2,1)m_c%
-40\,I_{6}^{[0,1]}(3,1,2)m_c-20\,I_{0}^{[0,1]}(3,1,1)m_b \\
&&-50\,I_{0}^{[0,1]}(2,1,2)m_b+80\,I_{6}^{[0,1]}(3,1,2)m_b%
+20\,I_{0}^{[0,1]}(3,1,2)m_b+40\,I_{6}^{[0,1]}(2,2,2)m_b \\
&&+30\,I_{{0}}(1,1,2)m_b-40\,I_{0}^{[0,1]}(1,2,2)m_b%
+80\,I_{6}^{[0,1]}(2,3,1)m_b+20\,\,I_{0}^{[0,2]}(2,3,1)m_b \\
&&+80\,I_{{6}}(3,1,1)m_b+30\,I_{0}^{[0,1]}(2,2,2)m_b%
+20\,I_{0}^{[0,1]}(1,3,1)m_b+80\,I_{{6}}(1,3,1)m_b \\
&&+40\,I_{{6}}(2,1,2)m_b+30\,I_{0}^{[0,2]}(3,2,1)m_b%
+40\,I_{6}^{[0,1]}(3,2,1)m_b-30\,I_{0}^{[0,1]}(2,2,1)m_b \\
&&-40\,I_{6}^{[0,2]}(3,2,2)m_b+30\,I_{{0}}(1,2,1)m_b+80\,I_{{%
6}}(2,2,1)m_b-30\,I_{{0}}(2,1,1)m_b \\
&&+40\,I_{{6}}(1,2,2)m_b   ~,
\end{eqnarray*}
\begin{eqnarray*}
&&C_{A_{2}}=10\,I_{{3}}(4,1,2){m_c}^{5}+10\,I_{{2}}(3,2,2){%
m_c}^{5}+20\,I_{{2}}(4,2,1){m_c}^{5}+10\,I_{{2}}(4,1,2){%
m_c}^{5} \\
&&-10\,I_{{5}}(4,1,2){m_c}^{5}+20\,I_{{1}}(4,2,1){m_c}%
^{5}+10\,I_{{0}}(3,2,2){m_c}^{5}+10\,I_{{1}}(4,1,2){m_c}^{5}
\\
&&+10\,I_{{1}}(3,2,2){m_c}^{5}+30\,I_{{5}}(3,3,1){m_c}%
^{5}-15\,I_{{1}}(3,3,1){m_c}^{5}-15\,I_{{2}}(3,3,1){m_c}^{5}
\\
&&-30\,I_{{5}}(4,2,1){m_c}^{5}-30\,I_{{3}}(3,3,1){m_c}%
^{5}+30\,I_{{3}}(4,2,1){m_c}^{5}-10\,I_{{3}}(3,2,2){m_c}^{4}%
m_b \\
&&+10\,I_{{1}}(4,2,1){m_c}^{4}m_b-10\,I_{{5}}(4,1,2){\mathit{%
m_c}}^{4}m_b-30\,I_{{5}}(3,3,1){m_c}^{4}m_b+80\,I_{{5}%
}(2,4,1){m_c}^{4}m_b \\
&&-10\,I_{{2}}(2,4,1){m_c}^{4}m_b-10\,I_{{1}}(2,4,1){\mathit{%
m_c}}^{4}m_b+10\,I_{{3}}(4,1,2){m_c}^{4}m_b+10\,I_{{2}%
}(4,2,1){m_c}^{4}m_b \\
&&+30\,I_{{3}}(3,3,1){m_c}^{4}m_b-80\,I_{{3}}(2,4,1){\mathit{%
m_c}}^{4}m_b+10\,I_{{5}}(3,2,2){m_c}^{4}m_b+40\,I_{{3}%
}(2,4,1){m_c}^{2}{m_b}^{3} \\
&&-40\,I_{{5}}(2,4,1){m_c}^{2}{m_b}^{3}-20\,I_{{3}}(2,4,1){%
m_b}^{5}+20\,I_{{5}}(2,4,1){m_b}^{5}+20\,I_{{1}}(2,2,2){%
m_c}^{3} \\
&&-30\,I_{{2}}(4,1,1){m_c}^{3}+10\,I_{2}^{[0,1]}(3,2,2){m_c}%
^{3}+20\,I_{{0}}(2,2,2){m_c}^{3}+20\,I_{{2}}(2,2,2){m_c}^{3}
\\
&&-15\,I_{{0}}(3,1,2){m_c}^{3}-10\,I_{1}^{[0,1]}(3,2,2){m_c}%
^{3}-30\,I_{{1}}(4,1,1){m_c}^{3}+40\,I_{{1}}(3,2,1){m_c}^{3}
\\
&&+20\,I_{{2}}(3,1,2){m_c}^{3}+40\,I_{{2}}(3,2,1){m_c}%
^{3}-30\,I_{{1}}(2,3,1){m_c}^{3}-30\,I_{{2}}(2,3,1){m_c}^{3}
\\
&&+20\,I_{{1}}(3,1,2){m_c}^{3}-20\,I_{{1}}(1,4,1){m_c}^{2}%
m_b-20\,I_{1}^{[0,1]}(2,4,1){m_c}^{2}m_b%
+20\,I_{2}^{[0,1]}(2,4,1){m_c}^{2}m_b \\
&&-5\,I_{{0}}(2,2,2){m_c}^{2}m_b+40\,I_{5}^{[0,1]}(2,4,1){%
m_c}^{2}m_b+20\,I_{{2}}(3,2,1){m_c}^{2}m_b%
+10\,I_{{0}}(3,1,2){m_c}^{2}m_b \\
&&-40\,I_{3}^{[0,1]}(2,4,1){m_c}^{2}m_b-20\,I_{{2}}(1,4,1){%
m_c}^{2}m_b+20\,I_{{1}}(3,2,1){m_c}^{2}m_b%
-5\,I_{1}^{[0,1]}(3,2,2)m_c\,{m_b}^{2} \\
&&-30\,I_{{1}}(1,4,1)m_c\,{m_b}^{2}+5\,I_{2}^{[0,1]}(3,2,2)%
m_c\,{m_b}^{2}-10\,I_{{0}}(2,2,2)m_c\,{m_b}%
^{2}-30\,I_{{2}}(1,4,1)m_c\,{m_b}^{2} \\
&&+5\,I_{{0}}(3,1,2)m_c\,{m_b}^{2}-40\,I_{5}^{[0,1]}(2,4,1){%
m_b}^{3}+40\,I_{3}^{[0,1]}(2,4,1){m_b}^{3}-5\,I_{{0}}(2,2,2){%
m_b}^{3} \\
&&-10\,I_{1}^{[0,1]}(2,2,2)m_c+15\,I_{{1}}(2,1,2)m_c+20\,I_{{%
2}}(2,2,1)m_c-15\,I_{2}^{[0,1]}(2,3,1)m_c \\
&&+15\,I_{{2}}(2,1,2)m_c+5\,I_{{0}}(2,1,2)m_c+5\,I_{{1}%
}(1,2,2)m_c-5\,I_{2}^{[0,2]}(3,2,2)m_c \\
&&+10\,I_{2}^{[0,1]}(3,2,1)m_c+10\,I_{{0}}(1,2,2)m_c-15\,I_{{%
2}}(3,1,1)m_c-15\,I_{{2}}(1,3,1)m_c \\
&&+5\,I_{{2}}(1,2,2)m_c+5\,I_{1}^{[0,2]}(3,2,2)m_c%
-5\,I_{0}^{[0,1]}(3,1,2)m_c+15\,I_{1}^{[0,1]}(2,3,1)m_c \\
&&-15\,I_{{1}}(3,1,1)m_c+20\,I_{{1}}(2,2,1)m_c-15\,I_{{1}%
}(1,3,1)m_c+10\,I_{2}^{[0,1]}(2,2,2)m_c \\
&&+10\,I_{2}^{[0,1]}(3,1,2)m_c-10\,I_{1}^{[0,1]}(3,2,1)m_c%
-10\,I_{1}^{[0,1]}(3,1,2)m_c+10\,I_{{2}}(2,2,1)m_b \\
&&-25\,I_{{0}}(2,1,2)m_b-10\,I_{2}^{[0,1]}(1,4,1)m_b%
+40\,I_{5}^{[0,1]}(1,4,1)m_b+5\,I_{2}^{[0,1]}(2,2,2)m_b \\
&&-5\,I_{{0}}(1,2,2)m_b+40\,I_{3}^{[0,1]}(2,3,1)m_b%
-40\,I_{5}^{[0,1]}(2,3,1)m_b+20\,I_{5}^{[0,2]}(2,4,1)m_b \\
&&+10\,I_{1}^{[0,1]}(1,4,1)m_b-5\,I_{1}^{[0,1]}(2,2,2)m_b%
+10\,I_{{1}}(2,2,1)m_b-40\,I_{3}^{[0,1]}(1,4,1)m_b \\
&&-20\,I_{3}^{[0,2]}(2,4,1)m_b+5\,I_{0}^{[0,1]}(2,2,2)m_b  ~,
\end{eqnarray*}
\begin{eqnarray*}
&&C_{V^{\prime }} =-20\,I_{{1}}(3,2,2){m_{c}}^{5}-20\,I_{{2}}(3,2,2){%
m_{c}}^{5}-20\,I_{{0}}(3,2,2){m_{c}}^{5}+20\,I_{{1}}(3,2,2){%
m_{c}}^{4}\mathit{m_b} \\
&&+20\,I_{{2}}(3,2,2){m_{c}}^{3}{\mathit{m_b}}^{2}+20\,I_{{1}}(3,2,2){%
m_{c}}^{3}{\mathit{m_b}}^{2}+20\,I_{{0}}(3,2,2){m_{c}}^{3}{%
\mathit{m_b}}^{2}-20\,I_{{1}}(3,2,2){m_{c}}^{2}{\mathit{m_b}}^{3} \\
&&-40\,I_{{0}}(2,2,2){m_{c}}^{3}+40\,I_{0}^{[0,1]}(3,2,2){m_{c}}%
^{3}-40\,I_{{2}}(2,2,2){m_{c}}^{3}-40\,I_{{1}}(2,2,2){m_{c}}^{3}
\\
&&+40\,I_{1}^{[0,1]}(3,2,2){m_{c}}^{3}-60\,I_{{0}}(4,1,1){m_{c}}%
^{3}+40\,I_{2}^{[0,1]}(3,2,2){m_{c}}^{3}-60\,I_{{2}}(4,1,1){m_{c}%
}^{3} \\
&&+20\,I_{{2}}(3,2,1){m_{c}}^{3}-20\,I_{{0}}(3,1,2){m_{c}}%
^{3}-60\,I_{{1}}(4,1,1){m_{c}}^{3}+60\,I_{{2}}(3,1,2){m_{c}}^{3}
\\
&&-20\,I_{{2}}(3,2,1){m_{c}}^{2}m_{b}-40\,I_{1}^{[0,1]}(3,2,2){%
m_{c}}^{2}m_{b}+20\,I_{{1}}(3,2,1){m_{c}}^{2}m_{b}%
+60\,I_{{1}}(4,1,1){m_{c}}^{2}m_{b} \\
&&+40\,I_{{1}}(2,3,1){{m_{c}}}^{2}m_{b}+80\,I_{{0}}(2,3,1){%
m_{c}}^{2}m_{b}+20\,I_{{0}}(3,2,1){m_{c}}^{2}m_{b}%
+40\,I_{{1}}(2,2,2){m_{c}}^{2}m_{b} \\
&&-60\,I_{{1}}(3,2,1)m_{c}\,{m_{b}}^{2}+20\,I_{1}^{[0,1]}(3,2,2)%
m_{c}\,{m_{b}}^{2}+20\,I_{0}^{[0,1]}(3,2,2)m_{c}\,{\mathit{%
m_b}}^{2}+120\,I_{{0}}(1,4,1)m_{c}\,{m_{b}}^{2} \\
&&-40\,I_{{0}}(3,2,1)m_{c}\,{m_{b}}^{2}-40\,I_{{2}}(3,2,1)%
m_{c}\,{m_{b}}^{2}+120\,I_{{2}}(1,4,1)m_{c}\,{m_{b}}%
^{2}+120\,I_{{1}}(1,4,1)m_{c}\,{m_{b}}^{2} \\
&&+20\,I_{2}^{[0,1]}(3,2,2)m_{c}\,{m_{b}}^{2}-120\,I_{{1}}(1,4,1)%
{m_{b}}^{3}-20\,I_{1}^{[0,1]}(3,2,2){m_{b}}^{3}+40\,I_{{1}%
}(3,2,1){m_{b}}^{3} \\
&&+20\,I_{{1}}(2,2,2){m_{b}}^{3}-40\,I_{{1}}(2,3,1){m_{b}}%
^{3}+60\,I_{0}^{[0,1]}(3,1,2)m_{c}+40\,I_{{0}}(2,2,1)m_{c} \\
&&-20\,I_{0}^{[0,2]}(3,2,2)m_{c}-20\,I_{{1}}(3,1,1)m_{c}-40\,I_{{%
1}}(1,2,2)m_{c}+20\,I_{{2}}(2,1,2)m_{c} \\
&&-40\,I_{{1}}(2,1,2)m_{c}+20\,I_{{2}}(3,1,1)m_{c}-20\,I_{{0}%
}(3,1,1)m_{c}+40\,I_{2}^{[0,1]}(2,2,2)m_{c} \\
&&+40\,I_{{1}}(2,2,1)m_{c}+40\,I_{0}^{[0,1]}(2,2,2)m_{c}-60\,I_{{%
0}}(2,1,2)m_{c}-20\,I_{2}^{[0,1]}(3,1,2)m_{c} \\
&&-20\,I_{1}^{[0,2]}(3,2,2)m_{c}+40\,I_{1}^{[0,1]}(3,2,1)m_{c}%
-40\,I_{{0}}(1,2,2m_{c}+20\,I_{2}^{[0,1]}(3,2,1)m_{c} \\
&&+60\,I_{{2}}(2,2,1)m_{c}+40\,I_{0}^{[0,1]}(3,2,1)m_{c}%
+40\,I_{1}^{[0,1]}(2,2,2)m_{c}-20\,I_{2}^{[0,2]}(3,2,2)m_{c} \\
&&+40\,I_{1}^{[0,1]}(3,1,2)m_{c}-40\,I_{{2}}(1,2,2)m_{c}+200\,I_{%
{0}}(1,3,1)m_{b}+80\,I_{2}^{[0,1]}(2,3,1)m_{b} \\
&&+40\,I_{{1}}(1,2,2)m_{b}+40\,I_{1}^{[0,1]}(2,3,1)m_{b}+60\,I_{{%
1}}(2,1,2)m_{b}-40\,I_{1}^{[0,1]}(3,1,2)m_{b} \\
&&-40\,I_{{0}}(2,2,1)m_{b}+120\,I_{{2}}(1,3,1)m_{b}-100\,I_{{1}%
}(2,2,1)m_{b}-40\,I_{1}^{[0,1]}(3,2,1)m_{b} \\
&&-40\,I_{1}^{[0,1]}(2,2,2)m_{b}+40\,I_{{1}}(1,3,1)m_{b}-40\,I_{{%
2}}(2,2,1)m_{b}+20\,I_{1}^{[0,2]}(3,2,2)m_{b}  ~,
\end{eqnarray*}
\begin{eqnarray*}
&&C_{A_{1}^{\prime }}=-10\,I_{{0}}(3,2,2){m_{c}}^{6}m_{b}+10\,I_{%
{0}}(3,2,2){m_{c}}^{5}{m_{b}}^{2}+10\,I_{{0}}(3,2,2){m_{c}}%
^{4}{m_{b}}^{3}-10\,I_{{0}}(3,2,2){m_{c}}^{3}{m_{b}}^{4} \\
&&-40\,I_{{6}}(3,2,2){m_{c}}^{5}+10\,I_{{0}}(2,1,3){m_{c}}^{4}%
m_{b}+40\,I_{{6}}(3,2,2){m_{c}}^{4}m_{b}-10\,I_{{0}}(3,2,1)%
{m_{c}}^{4}m_{b} \\
&&-30\,I_{{0}}(4,1,1){m_{c}}^{4}m_{b}-30\,I_{{0}}(2,2,2){\mathit{%
mc}}^{4}m_{b}+30\,I_{0}^{[0,1]}(3,2,2){m_{c}}^{4}m_{b}%
-20\,I_{{0}}(2,3,1){m_{c}}^{4}m_{b} \\
&&+40\,I_{{6}}(3,2,2){m_{c}}^{3}{m_{b}}^{2}+30\,I_{{0}}(4,1,1){%
m_{c}}^{3}{m_{b}}^{2}-10\,I_{{0}}(2,1,3){m_{c}}^{3}{%
m_{b}}^{2}+20\,I_{{0}}(2,2,2){m_{c}}^{3}{m_{b}}^{2} \\
&&-20\,I_{0}^{[0,1]}(3,2,2){m_{c}}^{3}{m_{b}}^{2}+10\,I_{{0}%
}(3,2,1){m_{c}}^{3}{m_{b}}^{2}-40\,I_{{6}}(3,2,2){m_{c}}%
^{2}{m_{b}}^{3}+60\,I_{{0}}(1,4,1){m_{c}}^{2}{m_{b}}^{3} \\
&&-20\,I_{{0}}(3,2,1){m_{c}}^{2}{m_{b}}^{3}+20\,I_{{0}}(2,3,1){%
m_{c}}^{2}{m_{b}}^{3}-60\,I_{{0}}(1,4,1)m_{c}\,{m_{b}%
}^{4}-10\,I_{0}^{[0,1]}(3,2,2)m_{c}\,{m_{b}}^{4} \\
&&+20\,I_{{0}}(3,2,1)m_{c}\,{m_{b}}^{4}-40\,I_{{6}}(3,1,2){%
m_{c}}^{3}+80\,I_{6}^{[0,1]}(3,2,2){m_{c}}^{3}+10\,I_{{0}}(2,1,2)%
{m_{c}}^{3} \\
&&-80\,I_{{6}}(2,2,2){m_{c}}^{3}+10\,I_{{0}}(3,1,1){m_{c}}%
^{3}-80\,I_{{6}}(3,2,1){m_{c}}^{3}-120\,I_{{6}}(4,1,1){m_{c}}^{3}
\\
&&-30\,I_{0}^{[0,2]}(3,2,2){m_{c}}^{2}m_{b}-80%
\,I_{6}^{[0,1]}(3,2,2){m_{c}}^{2}m_{b}+40\,I_{{6}}(3,1,2){%
m_{c}}^{2}m_{b}+30\,I_{{0}}(2,2,1){m_{c}}^{2}m_{b} \\
&&+20\,I_{0}^{[0,1]}(3,1,2){m_{c}}^{2}m_{b}+80\,I_{{6}}(3,2,1){%
m_{c}}^{2}m_{b}+40\,I_{0}^{[0,1]}(3,2,1){m_{c}}^{2}\mathit{%
m_b}+30\,I_{0}^{[0,1]}(4,1,1){m_{c}}^{2}m_{b} \\
&&+120\,I_{{6}}(4,1,1){m_{c}}^{2}m_{b}-20\,I_{0}^{[0,1]}(2,1,3){%
m_{c}}^{2}m_{b}+40\,I_{0}^{[0,1]}(2,3,1){m_{c}}^{2}\mathit{%
m_b}-50\,I_{{0}}(2,1,2){m_{c}}^{2}m_{b} \\
&&+20\,I_{{0}}(1,1,3){m_{c}}^{2}m_{b}+20\,I_{{0}}(1,3,1){\mathit{%
mc}}^{2}m_{b}-20\,I_{{0}}(3,1,1){m_{c}}^{2}m_{b}%
+60\,I_{0}^{[0,1]}(2,2,2){m_{c}}^{2}m_{b} \\
&&-40\,I_{{0}}(1,2,2){m_{c}}^{2}m_{b}-80\,I_{{6}}(2,3,1){\mathit{%
mc}}^{2}m_{b}+40\,I_{{6}}(2,2,2){m_{c}}^{2}m_{b}-40\,I_{{6}%
}(3,2,1)m_{c}\,{m_{b}}^{2} \\
&&+240\,I_{{6}}(1,4,1)m_{c}\,{m_{b}}^{2}+20\,I_{{0}}(1,2,2)%
m_{c}\,{m_{b}}^{2}-40\,I_{0}^{[0,1]}(3,1,2)m_{c}\,{\mathit{%
m_b}}^{2}+10\,I_{0}^{[0,1]}(2,1,3)m_{c}\,{m_{b}}^{2} \\
&&-20\,I_{0}^{[0,1]}(2,2,2)m_{c}\,{m_{b}}^{2}-10%
\,I_{0}^{[0,1]}(3,2,1)m_{c}\,{m_{b}}^{2}+40\,I_{6}^{[0,1]}(3,2,2)%
m_{c}\,{m_{b}}^{2}-60\,I_{{0}}(1,3,1)m_{c}\,{m_{b}}%
^{2} \\
&&-10\,I_{{0}}(1,1,3)m_{c}\,{m_{b}}^{2}+10\,I_{{0}}(3,1,1)%
m_{c}\,{m_{b}}^{2}+10\,I_{0}^{[0,2]}(3,2,2)m_{c}\,{\mathit{%
m_b}}^{2}-20\,I_{{0}}(2,2,1)m_{c}\,{m_{b}}^{2} \\
&&+40\,I_{{0}}(2,1,2)m_{c}\,{m_{b}}^{2}-40\,I_{{6}}(2,2,2)%
m_{c}\,{m_{b}}^{2}-10\,I_{{0}}(1,2,2){m_{b}}^{3}-40\,I_{{0}%
}(1,3,1){m_{b}}^{3} \\
&&+10\,I_{{0}}(2,2,1){m_{b}}^{3}+80\,I_{{6}}(2,3,1){m_{b}}%
^{3}+40\,I_{{6}}(3,2,1){m_{b}}^{3}-240\,I_{{6}}(1,4,1){m_{b}}^{3}
\\
&&+20\,I_{0}^{[0,1]}(2,3,1){m_{b}}^{3}+40\,I_{{6}}(2,2,2){m_{b}}%
^{3}-10\,I_{0}^{[0,1]}(3,2,2){m_{b}}^{3}+20\,I_{0}^{[0,1]}(2,2,2){%
m_{b}}^{3} \\
&&+30\,I_{0}^{[0,1]}(3,2,1){m_{b}}^{3}-60\,I_{0}^{[0,1]}(1,4,1){%
m_{b}}^{3}-40\,I_{6}^{[0,1]}(3,2,2){m_{b}}^{3}-80\,I_{{6}}(2,2,1)%
m_{c} \\
&&+20\,I_{{0}}(1,1,2)m_{c}-10\,I_{0}^{[0,1]}(2,1,2)m_{c}%
+10\,I_{0}^{[0,1]}(3,1,1)m_{c}+80\,I_{{6}}(3,1,1)m_{c} \\
&&+40\,I_{{6}}(2,1,2)m_{c}+40\,I_{6}^{[0,1]}(3,1,2)m_{c}%
-40\,I_{6}^{[0,2]}(3,2,2)m_{c}-20\,I_{{0}}(2,1,1)m_{c} \\
&&+80\,I_{6}^{[0,1]}(3,2,1)m_{c}+10\,I_{{0}}(1,2,1)m_{c}+80\,I_{{%
6}}^{[0,1]}(2,2,2)m_{c}-40\,I_{6}^{[0,1]}(2,2,2)m_{b} \\
&&-80\,I_{6}^{[0,1]}(3,1,2)m_{b}-20\,I_{0}^{[0,2]}(2,3,1)m_{b}%
+30\,I_{0}^{[0,1]}(2,1,2)m_{b}-30\,I_{{0}}(1,2,1)m_{b} \\
&&+40\,I_{6}^{[0,2]}(3,2,2)m_{b}+40\,I_{{6}}(2,2,1)m_{b}-40\,I_{{%
6}}(2,1,2)m_{b}+40\,I_{{0}}(2,1,1)m_{b} \\
&&+10\,I_{0}^{[0,2]}(2,1,3)m_{b}-20\,I_{0}^{[0,1]}(1,3,1)m_{b}%
+20\,I_{0}^{[0,1]}(3,1,1)m_{b}-20\,I_{0}^{[0,1]}(1,1,3)m_{b} \\
&&-30\,I_{0}^{[0,2]}(2,2,2)m_{b}+40\,I_{0}^{[0,1]}(1,2,2)m_{b}%
-80\,I_{{6}}(3,1,1)m_{b}+30\,I_{0}^{[0,1]}(2,2,1)m_{b} \\
&&-20\,I_{0}^{[0,2]}(3,1,2)m_{b}-40\,I_{{6}}(1,2,2)m_{b}%
-40\,I_{6}^{[0,1]}(3,2,1)m_{b}-80\,I_{6}^{[0,1]}(2,3,1)m_{b} \\
&&-80\,I_{{6}}(1,3,1)m_{b}-30\,I_{0}^{[0,2]}(3,2,1)m_{b}-30\,I_{{%
0}}(1,1,2)m_{b}  ~,
\end{eqnarray*}
\begin{eqnarray*}
&&C_{A_{2}^{\prime }}=-10\,I_{{1}}(3,2,2){m_{c}}^{5}-10\,I_{{3}}(3,2,2)%
{m_{c}}^{5}-5\,I_{{0}}(3,2,2){m_{c}}^{5}+10\,I_{{5}}(3,2,2){%
m_{c}}^{5} \\
&&-5\,I_{{2}}(3,2,2){m_{c}}^{4}m_{b}+5\,I_{{1}}(3,2,2){\mathit{mc%
}}^{4}m_{b}-10\,I_{{5}}(3,2,2){m_{c}}^{4}m_{b}+10\,I_{{3}%
}(3,2,2){m_{c}}^{4}m_{b} \\
&&+10\,I_{{3}}(3,2,2){m_{c}}^{3}{m_{b}}^{2}+10\,I_{{1}}(3,2,2){%
m_{c}}^{3}{m_{b}}^{2}+5\,I_{{0}}(3,2,2){m_{c}}^{3}{\mathit{%
m_b}}^{2}-10\,I_{{5}}(3,2,2){m_{c}}^{3}{m_{b}}^{2} \\
&&+10\,I_{{5}}(3,2,2){m_{c}}^{2}{m_{b}}^{3}-10\,I_{{3}}(3,2,2){%
m_{c}}^{2}{m_{b}}^{3}-5\,I_{{1}}(3,2,2){m_{c}}^{2}{\mathit{%
mb}}^{3}+5\,I_{{2}}(3,2,2){m_{c}}^{2}{m_{b}}^{3} \\
&&+30\,I_{1}^{[0,1]}(3,2,2){m_{c}}^{3}-10\,I_{{0}}(2,2,2){m_{c}}%
^{3}+10\,I_{{2}}(3,1,2){m_{c}}^{3}-20\,I_{{1}}(3,1,2){m_{c}}^{3}
\\
&&-5\,I_{{0}}(3,1,2){m_{c}}^{3}+10\,I_{0}^{[0,1]}(3,2,2){m_{c}}%
^{3}-25\,I_{{1}}(3,2,1){m_{c}}^{3}+20\,I_{3}^{[0,1]}(3,2,2){m_{c}%
}^{3} \\
&&-15\,I_{{2}}(3,2,1){m_{c}}^{3}-25\,I_{{0}}(3,2,1){m_{c}}%
^{3}-20\,I_{5}^{[0,1]}(3,2,2){m_{c}}^{3}-15\,I_{{0}}(4,1,1){m_{c}%
}^{3} \\
&&-10\,I_{2}^{[0,1]}(3,2,2){m_{c}}^{3}-30\,I_{{1}}(4,1,1){m_{c}}%
^{3}-20\,I_{{1}}(2,2,2){m_{c}}^{3}+5\,I_{{1}}(2,2,2){m_{c}}^{2}%
m_{b} \\
&&-10\,I_{{1}}(2,3,1){m_{c}}^{2}m_{b}-5\,I_{{2}}(2,2,2){\mathit{%
mc}}^{2}m_{b}-20\,I_{{0}}(2,3,1){m_{c}}^{2}m_{b}%
+20\,I_{5}^{[0,1]}(3,2,2){m_{c}}^{2}m_{b} \\
&&-10\,I_{1}^{[0,1]}(3,2,2){m_{c}}^{2}m_{b}-20%
\,I_{3}^{[0,1]}(3,2,2){m_{c}}^{2}m_{b}+10\,I_{{2}}(2,3,1){%
m_{c}}^{2}m_{b}+10\,I_{{1}}(3,1,2){m_{c}}^{2}m_{b} \\
&&+15\,I_{{1}}(4,1,1){m_{c}}^{2}m_{b}+10\,I_{2}^{[0,1]}(3,2,2){%
m_{c}}^{2}m_{b}+5\,I_{{0}}(3,2,1){m_{c}}^{2}m_{b}%
+20\,I_{{1}}(3,2,1){m_{c}}^{2}m_{b} \\
&&-15\,I_{{2}}(4,1,1){m_{c}}^{2}m_{b}-10\,I_{{2}}(3,1,2){\mathit{%
mc}}^{2}m_{b}+10\,I_{{2}}(3,2,1){m_{c}}^{2}m_{b}+10\,I_{{2}%
}(2,2,2)m_{c}\,{m_{b}}^{2} \\
&&-20\,I_{{1}}(3,2,1)m_{c}\,{m_{b}}^{2}+30\,I_{{0}}(1,4,1)%
m_{c}\,{m_{b}}^{2}-10\,I_{{1}}(2,2,2)m_{c}\,{m_{b}}%
^{2}+10\,I_{3}^{[0,1]}(3,2,2)m_{c}\,{m_{b}}^{2} \\
&&+5\,I_{0}^{[0,1]}(3,2,2)m_{c}\,{m_{b}}^{2}-15\,I_{{0}}(3,2,1)%
m_{c}\,{m_{b}}^{2}-5\,I_{2}^{[0,1]}(3,2,2)m_{c}\,{\mathit{%
m_b}}^{2}+15\,I_{1}^{[0,1]}(3,2,2)m_{c}\,{m_{b}}^{2} \\
&&+60\,I_{{1}}(1,4,1)m_{c}\,{m_{b}}^{2}-5\,I_{{2}}(3,1,2)\mathit{%
mc}\,{m_{b}}^{2}-10\,I_{5}^{[0,1]}(3,2,2)m_{c}\,{m_{b}}%
^{2}+5\,I_{{1}}(3,1,2)m_{c}\,{m_{b}}^{2} \\
&&-10\,I_{{2}}(3,2,1){m_{b}}^{3}+5\,I_{2}^{[0,1]}(3,2,2){m_{b}}%
^{3}-30\,I_{{1}}(1,4,1){m_{b}}^{3}+10\,I_{{1}}(2,3,1){m_{b}}^{3}
\\
&&-5\,I_{1}^{[0,1]}(3,2,2){m_{b}}^{3}+10\,I_{{1}}(3,2,1){m_{b}}%
^{3}+10\,I_{5}^{[0,1]}(3,2,2){m_{b}}^{3}-10\,I_{3}^{[0,1]}(3,2,2){%
m_{b}}^{3} \\
&&+30\,I_{{2}}(1,4,1){m_{b}}^{3}-10\,I_{{2}}(2,3,1){m_{b}}%
^{3}+20\,I_{3}^{[0,1]}(2,2,2)m_{c}-20\,I_{{0}}(2,2,1)m_{c} \\
&&-5\,I_{0}^{[0,2]}(3,2,2)m_{c}-20\,I_{5}^{[0,1]}(3,2,1)m_{c}%
-15\,I_{1}^{[0,2]}(3,2,2)m_{c}+10\,I_{3}^{[0,1]}(3,1,2)m_{c} \\
&&-10\,I_{3}^{[0,2]}(3,2,2)m_{c}+5\,I_{{2}}(3,1,1)m_{c}-10\,I_{{0%
}}(3,1,1)m_{c}+10\,I_{5}^{[0,2]}(3,2,2)m_{c} \\
&&+30\,I_{1}^{[0,1]}(3,1,2)m_{c}+30\,I_{1}^{[0,1]}(2,2,2)m_{c}%
+20\,I_{3}^{[0,1]}(3,2,1)m_{c}+15\,I_{0}^{[0,1]}(3,2,1)m_{c} \\
&&-25\,I_{{1}}(2,2,1)m_{c}-20\,I_{5}^{[0,1]}(2,2,2)m_{c}-15\,I_{{%
0}}(2,1,2)m_{c}-20\,I_{{1}}(2,1,2)m_{c} \\
&&-10\,I_{5}^{[0,1]}(3,1,2)m_{c}+15\,I_{0}^{[0,1]}(3,1,2)m_{c}%
+5\,I_{2}^{[0,2]}(3,2,2)m_{c}+5\,I_{{1}}(3,1,1)m_{c} \\
&&-15\,I_{2}^{[0,1]}(3,2,1)m_{c}-10\,I_{2}^{[0,1]}(2,2,2)m_{c}%
-10\,I_{{2}}(2,1,2)m_{c}+35\,I_{1}^{[0,1]}(3,2,1)m_{c} \\
&&-5\,I_{{2}}(2,2,1)m_{c}+10\,I_{0}^{[0,1]}(2,2,2)m_{c}%
-10\,I_{1}^{[0,1]}(2,3,1)m_{b}+10\,I_{{0}}(1,3,1)m_{b} \\
&&+5\,I_{1}^{[0,2]}(3,2,2)m_{b}-10\,I_{3}^{[0,1]}(2,2,2)m_{b}%
+10\,I_{2}^{[0,1]}(3,2,1)m_{b}+15\,I_{{2}}(1,2,2)m_{b} \\
&&-20\,I_{{2}}(2,2,1)m_{b}+10\,I_{3}^{[0,2]}(3,2,2)m_{b}-15\,I_{{%
1}}(1,2,2)m_{b}+20\,I_{5}^{[0,1]}(3,1,2)m_{b} \\
&&-5\,I_{2}^{[0,2]}(3,2,2)m_{b}+20\,I_{{2}}(2,1,2)m_{b}%
+10\,I_{2}^{[0,1]}(2,3,1)m_{b}-20\,I_{3}^{[0,1]}(3,1,2)m_{b} \\
&&-20\,I_{3}^{[0,1]}(2,3,1)m_{b}+10\,I_{5}^{[0,1]}(2,2,2)m_{b}%
+20\,I_{5}^{[0,1]}(2,3,1)m_{b}+10\,I_{2}^{[0,1]}(3,1,2)m_{b} \\
&&-10\,I_{1}^{[0,1]}(3,1,2)m_{b}+10\,I_{5}^{[0,1]}(3,2,1)m_{b}%
-10\,I_{{0}}(2,2,1)m_{b}-10\,I_{1}^{[0,1]}(3,2,1)m_{b}
\end{eqnarray*}

where \bea \hat{I}_n^{[i,j]} (a,b,c) = \ga M_1^2 \dr^i \ga M_2^2
\dr^j \frac{d^i}{d\ga M_1^2 \dr^i} \frac{d^j}{d\ga M_2^2 \dr^j}
\left[\ga M_1^2 \dr^i \ga M_2^2 \dr^j \hat{I}_n(a,b,c) \right]~.
\nnb \eea

\end{document}